\documentclass[useAMS,usenatbib,aas_macros]{mn2e}

\usepackage{epsfig}
\usepackage{lscape,graphicx,colortbl}
\usepackage{amssymb}
\usepackage{natbib}
\usepackage{times}
\usepackage{aas_macros}

\newcommand{\msun}{M_\odot}
\newcommand{\ifm}[1]{\relax\ifmmode#1\else$\mathsurround=0pt#1$\fi}
\newcommand{\kms}{\ifmmode\,{\rm km}\,{\rm s}^{-1}\else km$\,$s$^{-1}$\fi}

\newcommand{\hmsun}{\,\ifm{h^{-1}}{M_{\odot}}}
\def\omm{\Omega_{\rm m}}
\def\oml{\Omega_{\Lambda}}

\newcommand{\dd}{{\rm d}}

\newcommand{\be}{\begin{equation}}
\newcommand{\ee}{\end{equation}}
\newcommand{\bea}{\begin{eqnarray}}
\newcommand{\eea}{\end{eqnarray}}
\newcommand{\z}{\emph{z}}
\newcommand{\mfal}{M_{\rm infall}}
\newcommand{\mfala}{M_{\rm infall}^1}
\newcommand{\mfalb}{M_{\rm infall}^2}
\newcommand{\ms}{m_{\star}}
\newcommand{\msi}{m_{\star,i}}
\newcommand{\msh}{m_{\star}-M_{\rm infall}}
\newcommand{\mf}{M_{200}}
\newcommand{\mfa}{M_{200}^1}
\newcommand{\mfb}{M_{200}^2}
\newcommand{\fof}{{\scshape fof~}}
\def\zfal{\z_{\rm infall}}
\def\adf{\alpha_{\rm df}}
\newcommand{\mfalt}{\widetilde{M}_{\rm infall}}
\newcommand{\mft}{\widetilde{M}_{200}}
\newcommand{\mfalo}{\overline{M}_{\rm infall}}
\newcommand{\mfo}{\overline{M}_{200}}

\begin{document}

\title[The stellar mass of central and satellite galaxies]
      {A tale of two populations: the stellar mass of central and satellite galaxies}

\author[E. Neistein et al.]
{Eyal Neistein$^{1}$\thanks{E-mail:$\;$eyal@mpe.mpg.de},
Cheng Li$^{2}$, Sadegh Khochfar$^{1}$, Simone M. Weinmann$^{3}$,
\newauthor Francesco Shankar$^{4}$, Michael Boylan-Kolchin$^{5}$
\\ \\
$^{1}$ Max-Planck-Institute for Extraterrestrial Physics, Giessenbachstrasse 1, 85748 Garching, Germany\\
$^{2}$ Max-Planck-Institute Partner Group, Key Laboratory for Research in
Galaxies and Cosmology, \\ \quad \quad Shanghai Astronomical Observatory, Chinese
Academy of Sciences, Nandan Road 80, Shanghai 200030, China \\
$^{3}$ Leiden Observatory, Leiden University, P.O. Box 9513, 2300 RA Leiden, The Netherlands \\
$^{4}$Max-Planck-Institute for Astrophysics, Karl-Schwarzschild-Str. 1, 85748 Garching, Germany \\
$^{5}$Center for Galaxy Evolution, 4129 Reines Hall, University of California, Irvine, CA 92697, USA
\\
\\}


\date{}
\pagerange{\pageref{firstpage}--\pageref{lastpage}} \pubyear{2011}
\maketitle

\label{firstpage}


\begin{abstract}
We develop a new empirical methodology to study the relation between the stellar mass of
galaxies and the mass of their host subhaloes. Our approach is similar to abundance
matching, and is based on assigning a stellar mass to each subhalo within a large
cosmological $N$-body simulation. However, we significantly extend the method to include
a different treatment for central and satellite galaxies, allowing the stellar mass of satellite
galaxies to depend on both the host subhalo mass, and on the halo mass.
Unlike in previous studies, our models are constrained by two observations: the stellar
mass function of galaxies, and their auto-correlation functions (CFs).
We present results for $\sim10^6$ different successful models, showing that
the uncertainty in the host subhalo mass reaches a factor of $\sim$10 for a given stellar mass.
Our results thus indicate that the host subhalo mass of central and satellite galaxies is
poorly constrained by using information coming solely from the abundance and CFs of
galaxies.
In addition, we demonstrate that the fraction of stellar mass relative to the universal
baryon fraction can vary between different models, and can reach $\sim0.6$ for
a specific population of low mass galaxies.
We conclude that additional observational constraints are needed,
in order to better constrain the mass relation between haloes and galaxies. These might include
weak lensing, satellite kinematics, or measures of clustering other than the CFs.
Moreover, CFs based on larger surveys with a better estimate of the systematic uncertainties are needed.
\end{abstract}




\section{Introduction}
\label{sec:intro}

The relation between the stellar mass of galaxies and their host dark-matter haloes
has become a key point of reference for many different theoretical and observational studies.
It summarizes in a simple way the complexity of galaxy formation physics when evolved
within growing dark-matter structure. A special attention was given
to the relation between galaxies and their host \emph{subhaloes}, which are the sub-structure bound density
peaks inside a halo.

The `abundance matching' (hereafter ABM) methodology is an important theoretical tool for
constraining the mass relation between galaxies and their host subhaloes
\citep{Kravtsov04,Vale04,Conroy06,Shankar06,Guo10a,Moster10,Behroozi10}.
In this simple empirical approach one assigns a model galaxy to each subhalo within a
cosmological $N$-body simulation. Assuming there is a one-to-one, monotonic
relation between the stellar mass ($\ms$) and the subhalo mass
($\mfal$, see section \ref{sec:subhalos}), the abundance of galaxies and subhaloes can be matched,
yielding a unique relation between $\ms$ and $\mfal$.
Surprisingly, this simple model provides a good fit to the observed clustering
properties of galaxies.

The ABM approach thus offers a practical solution to the relation between
subhaloes and galaxies, without going into the complex details of
galaxy formation physics. It can be used to constrain the mass
relation at various redshifts, to predict the star-formation rate in
galaxies \citep{Conroy09}, to study the
merger-rates of galaxies \citep{Hopkins10}, and to interpret
high-resolution hydrodynamical simulations \citep{Sawala11}.
In comparison to models based on the halo occupation distribution
\citep[][and references therein]{Berlind02,Cooray02,Tinker05,Zehavi05},
ABM uses explicit information on the location and mass of subhaloes,
decreasing the number of free parameters needed in the model.

The success of ABM is intriguing and raises several interesting questions:
Is it based on the only possible set of assumptions that can reproduce
the abundance and clustering of galaxies?
Do we miss models that result in a different $\msh$ relation? What are the
important assumptions made by ABM? How can we explore these
assumptions and test to what level they are constrained by observations?
In this work we try to address these questions. We specifically focus our results
on the freedom in the $\msh$ relation.

In ABM, the stellar mass is assigned to each subhalo according to $\mfal$.
For a satellite subhalo\footnote{We define `satellite subhaloes' as all the substructure clumps
within a \fof group, except the most massive one.}, $\mfal$ is defined as the mass at the last time
it was the most massive substructure within its \fof group. This is
a reasonable assumption because the subhalo mass of satellite galaxies can be
significantly stripped after falling into a larger dark-matter halo \citep[e.g.][]{Zentner05,vdBosch05},
more so than the stellar mass of its galaxy \citep[for stellar stripping see][]{Monaco06,Purcell07,Conroy07a,Yang09a}.
On the other hand, for a central subhalo, $\mfal$ is defined as its current mass.
\citet{Wang06} found that the relation between $\mfal$ and the stellar mass
of galaxies is tight in semi-analytic models, justifying the above
definition of $\mfal$.

There are various other assumptions made by ABM, which are mainly related to the treatment of
satellite galaxies. In a recent paper \citep[][hereafter paper I]{Neistein11}, we have
examined these assumptions using the semi-analytic model (SAM) of \citet[][]{Neistein10}.
Although ABM models assume that for a given $\mfal$, the stellar mass of central and satellite galaxies
is the same, there are various effects within SAMs that violate this assumption.
First, the relation between stellar mass and subhalo mass evolves with
redshift for central galaxies, affecting satellite galaxies at the time of infall \citep[the
typical infall time is a few Gyrs ago;][]{Wang07}.
Second, the stellar mass of satellite galaxies might be
different already at the time of infall from that of central galaxies at the same time. This is
because galaxies that join larger groups as satellites are located
in large-scale environment of higher density than the average. Consequently, the
dark-matter merger-histories of these galaxies are already
different at early epochs \citep[see also][]{Gao05,Harker06}.
Third, once a galaxy becomes a satellite, its stellar mass might
still grow. This is especially true for models in which
gas stripping in satellite galaxies is modeled
on time-scales of a few Gyrs \citep{Weinmann10}.
All the effects above are consistent with various studies about the properties
of central and satellite galaxies
\citep{Weinmann06,vdLinden07,Khochfar08,Skibba09,Pasquali09,Skibba11}.

In paper I, we showed that the effects above can change not only
the $\msh$ relation for satellite galaxies, but also the auto-correlation function (CF) of
galaxies.
If satellite and central galaxies with a given $\mfal$ are randomly redistributed in a SAM, the CFs can vary by
up to a factor of four. Even when just redistributing satellite and central galaxies among themselves, the
modifications in the CFs can reach a factor of 2 for massive galaxies.
We have shown in paper I that the CFs of SAM
galaxies can be reproduced accurately by the ABM approach only when the
stellar mass of satellite galaxies is assumed to depend on both
$\mfal$ and the host halo mass at $\z=0$. This finding is very
useful as it saves us the complex modeling of $\ms$ as a function of
the various effects mentioned above.

In this paper we
make use of the conclusions made in paper I, and add more ingredients to an
ABM model. We would like to study the influence of the new ingredients on the relation
between $\mfal$ and $\ms$, while the models are constrained to fit the observational data.
We allow the stellar mass of satellite galaxies to depend on both the $\mfal$ and
the host halo mass, and to deviate from the behaviour of central galaxies.
In addition, we use two ingredients for modeling the location
and abundance of satellite galaxies: Subhaloes
that were lost by the cosmological simulation at high
redshift, but can still host galaxies at $\z=0$, are identified
using an estimate for dynamical friction. The location of unresolved
subhaloes is fixed using either the location of the most bound particle of the
last identified subhalo, or using an analytical model with a free parametrization.

Once the population of galaxies is broken into two sub-populations,
the model cannot be constrained by the abundance of galaxies alone. The
number of models that can fit the observed stellar mass function of
galaxies is infinite. We therefore check a very large number of models ($\sim10^{12}$)
and look for those that fit both the observed stellar mass function,
and the observed auto-correlation function of galaxies. We will show
that even when using these two constraints, there is a
significant amount of freedom in the relation between $\ms$ and
$\mfal$. Satellite galaxies might occupy a significantly lower value
of $\mfal$ and thus might be more abundant than central galaxies at a fixed $\ms$. The
dependence of satellite galaxies on the halo mass is able to compensate for
this effect in terms of clustering, and is crucial for fitting the observed auto-correlation
functions.

The observational constraints used in this work include the stellar mass function of galaxies, as
derived by \citet{Li09}. For the CFs of galaxies we use the
methodology presented by \citet{Li06},
using the same stellar masses as in \citet{Li09}.
These stellar masses are based on redshift and 
five-band photometric data, as described in detail by \citet{Blanton07}. The galaxy sample is based
on the final data release \citep[DR7;][]{Abazajian09} of the Sloan Digital Sky Survey
\citep[SDSS;][]{York00}. Although there might be both random and systematic
deviations in the stellar masses used here, the two
observational constraints are self-consistent.

This paper is organized as follows. Section \ref{sec:formalism} describes in detail
the approach we use in this work. We elaborate on the new ingredients used here,
and the way we implement the models. In section \ref{sec:examples} we demonstrate the
results of our formalism by showing a few different models
that fit the data well. Our method for scanning the parameter space is discussed in section
\ref{sec:results}, where we explain how the parameter space is sampled, and how we
select good models. In this section we also show the properties of the successful
models. Lastly, we summarize our results and discuss them in section \ref{sec:discuss}.


\section{The formalism}
\label{sec:formalism}

In this section we describe the formalism developed for modeling the abundance and
clustering of galaxies. In general, we assume that the stellar mass of galaxies depends
solely on the properties of the host dark-matter haloes. This assumption is similar to
the abundance matching approach \citep[e.g.][]{Vale04}. However, we significantly extend
the ingredients of the model with respect to previous studies. This is
motivated by the analysis done in paper I, based on the semi-analytic models of
\citet{Neistein10}. The additional ingredients
are related to the assumptions regarding satellite galaxies: their abundance,
location, and the effect of the host halo mass on their stellar mass.

We start by summarizing the various ingredients of the formalism, a detailed description of
each component is given in the following sub-sections.
\begin{enumerate}
  \item All the subhaloes from a cosmological $N$-body simulation at $\z=0$ are selected.
  \item In addition, we use subhaloes that have been merged into a
  bigger structure at high redshift, but might host galaxies
  that survive until $\z=0$. This is due to the effects of dynamical 
friction and stripping on galaxies, which are not included in the $N$-body simulation.
  \item We use a freely tunable prefactor in the dynamical friction formula.
  \item We assign one galaxy with a specific stellar mass ($\ms$) to each subhalo from
  the full sample defined in items (i) \& (ii).
  \item For central subhaloes within their \fof group, we assume
  that the stellar mass of galaxies depends on the host subhalo mass at $\z=0$
  only ($\mfal$).
  \item For satellite subhaloes, we assume
  that the stellar mass of galaxies depends on two parameters: the host subhalo mass at
  the time of infall, and the host halo mass at $\z=0$.
  \item We allow the dependence of stellar mass on the subhalo mass to be different
  for central and satellite subhaloes.
  \item The location of subhaloes that are not found in the
  simulation at $\z=0$ is set by either the location of their most
  bound particle, or by an analytical model.
\end{enumerate}
Items (iii), (vi), (vii), (viii) in the list are new with respect to
previous ABM models, and their influence on clustering and abundance
of galaxies has not been tested before.


\subsection{Subhaloes and merger-trees}
\label{sec:subhalos}

We use merger trees extracted from the Millennium $N$-body simulation \citep{Springel05}.
This simulation was run using the cosmological parameters $(\omm,\,\oml,\,h,\,\sigma_8)=
(0.25,\,0.75,\,0.73,\,0.9)$, with a particle mass of $8.6\times10^8\,\hmsun$ and a box size
of 500 $h^{-1}$Mpc. The merger trees are based on snapshots spaced by $\sim$250 Myr, linking
\emph{subhaloes} identified by the \textsc{subfind} algorithm \citep{Springel01}. 
Subhaloes are defined as the bound
density peaks inside \fof groups \citep{Davis85}. More details on the simulation and the
subhalo merger-trees can be found in \citet{Springel05} and \citet{Croton06}.

The subhalo mass, $M_h$, corresponds to the number of particles inside a subhalo,
as identified by \textsc{subfind}. The \emph{infall} mass of the subhalo,
$\mfal$, is defined as
\begin{equation}
\label{eq:minfall}
\mfal = \left\{ \begin{array}{ll}
M_h & \textrm{if central within its \fof group}\\ \;\;\;\
& \;\;\;\;\;\;\;\;\;\;\;\;\;\;\;\;\;\;\;\;\;\;\;\;\;\;\;\;\;\;\;\;\;\;\;  \\
M_{h,p}(\zfal) & \textrm{otherwise}
\end{array} \right.
\end{equation}
Here $\zfal$ is the lowest redshift at which the main progenitor\footnote{
Main-progenitor histories are derived by following back in time the most massive
progenitor in each merger event.}
of the subhalo $M_h$ was the most massive within its \fof group, and $M_{h,p}$ is
the main progenitor mass at this redshift.

In addition to $\mfal$ we will use the \emph{halo} mass
$\mf$, which is defined as the mass within the radius where the halo has an
over-density 200 times the critical density of the simulation.
In general, the mass $\mf$ should include both the central subhalo within a group,
and all of its satellite subhaloes. However, due to the spherical symmetry forced on
$\mf$ and the specific over density being used here, it might deviate from the
exact \fof group mass. In what follows, we will refer to $\mf$ as the `halo mass' to
indicate that this mass is computed over a larger spatial region than the subhalo mass.
For satellite subhaloes, we assign the same value of $\mf$ as
is computed for their central subhalo
within the same group. Due to the effects above, the value of
$\mfal$ is often similar or higher than $\mf$ for central subhaloes\footnote{
For central subhaloes, $\mf$ is slightly smaller than $\mfal$. The average
difference is $\sim0.08$ dex, with an RMS scatter of $\sim0.06$ dex}.


\subsection{Satellite subhaloes: dynamical friction and location}

Satellite subhaloes are defined as all subhaloes inside
a \fof group except the most massive (central) subhalo. Within the $N$-body
simulation, satellite subhaloes lose their mass
while falling into a bigger subhalo, and thus might fall below the resolution limit
used in the \textsc{subfind} algorithm (20 particles for the Millennium simulation used here).
However, the galaxies residing inside these satellite subhaloes might live
longer, as they are more dense, and thus less vulnerable to
stripping. This effect is significant even for relatively high resolution
cosmological simulations, as was shown in paper I. It can
modify the abundance of subhaloes even at two orders of magnitude above
the minimum subhalo mass resolved by the simulation.

In order to take this effect into account, we model the time it takes the galaxy to fall into
the central galaxy by dynamical friction. At the last time the satellite subhalo
is resolved we compute its distance from the central  subhalo ($r_{\rm sat}$),
and estimate the dynamical friction time using the formula of \citet{Binney87},
\begin{equation}
t_{\rm df} = \adf \cdot \, \frac{1.17 V_v r_{\rm sat}^2}{G M_{h,2}\ln\left(
1+ M_{h,1}/M_{h,2} \right) } \, .
\label{eq:t_df}
\end{equation}
Here $M_{h,1}$ is the mass of the central subhalo, $V_v$ is its virial
velocity, and
$M_{h,2}$ is the mass of the satellite subhalo.
Once a satellite subhalo falls together with its central
subhalo into a larger group, we update $t_{\rm df}$ for both
objects according to the new central subhalo.

The dynamical friction estimate is computed using the mass of subhaloes only, before galaxies
are being modeled. Therefore, the formula deviates from its proper definition,
as it does not include the effect of the galaxy mass on dynamical friction. However,
since this formula uses various simplified assumptions with a larger
uncertainty \citep[for example,
the exact trajectories of the satellite subhaloes are ignored, see][]{Boylan08},
one general constant, $\adf$, is being used here to absorb all the related
uncertainties. Semi-analytic models often use
$\adf\sim 2-3$ \citep{DeLucia07,Khochfar09,Neistein10},
in agreement with more detailed studies of dynamical friction processes
\citep{Colpi99,Boylan08,Jiang08,Mo10}. Since we ignore contribution due to the galaxy 
mass, we allow more freedom in $\adf$ as is discussed below.

To summarize, three types of subhaloes exist in our models:
\begin{itemize}
\item {\bf central subhaloes:} most massive subhaloes within their \fof group.
\item {\bf satellite subhaloes:} all subhaloes except central subhaloes.
\item {\bf unresolved subhaloes:} subhaloes that were last identified at high
$\z$, and are added according to the dynamical friction formula
above. All the unresolved subhaloes are also satellite subhaloes.
\end{itemize}
The location of unresolved subhaloes is not given by the simulation. We therefore
need to estimate the location of the galaxies that are assigned to these subhaloes.
This issue was treated by e.g. \citet{Croton06} who used the location
of the most bound particle of the last identified subhalo, as is given by the dark-matter
simulation at $\z=0$. The same location is adopted here as well. However, we would like
to take into account the uncertainty involved in this method. Obviously,
one dark-matter particle inside a collisionless $N$-body simulation cannot
mimic accurately the location of an extended galaxy, with a possibly very different
mass. We therefore use the following analytical model as an additional option:
\begin{equation}
r= r_{\rm sat} \left( 1-\tau^p \right) ^{1/q} \,.
\label{eq:sat_loc}
\end{equation}
Here $\tau$ is the fraction of time spent out of all the estimated dynamical friction time
until $\z=0$, $r$ is the distance we adopt at $\z=0$ from the central subhalo, and $p,q$
are constants. In appendix \ref{sec:sat_loc} we derive this model by following the angular momentum of a particle
inside a spherical gravitational potential of the type $\rho\propto r^{-\gamma}$. We
show that for a specific values of $p$ and $q$, the model is able to reproduce the location of subhaloes
as is modeled by the most bound particle. This parametrization
enables us to check various modifications to the location of unresolved subhaloes.


\subsection{Counting subhaloes}

\begin{figure}
\centerline{ \hbox{ \epsfig{file=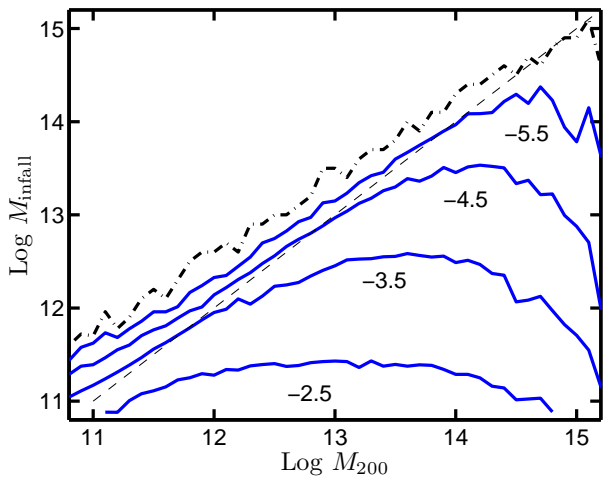,width=9cm} }}
\caption{The abundance of satellite subhaloes, $\phi_s$, as a function of $\mfal$ and
$\mf$, using the full Millennium simulation with $\adf=3$. \emph{Contours} show constant values
of $\phi_s$ as labeled in units of Log Mpc$^{-3}$ dex$^{-2}$. 
$\phi_s$ is bigger than zero only below the \emph{dotted-dashed} line. 
The \emph{thin dashed} line corresponds to $\mfal=\mf$ and is shown as a reference.}
  \label{fig:nsat_2d}
\end{figure}

The usual way to implement ABM is to first populate subhaloes
with galaxies, and only then to compute the auto correlation functions (CFs) of
galaxies. In our approach there are many possible models, mainly
depending on the different mass relations for satellite and central
galaxies. Since computing the statistics of pairs is the most demanding
computational step, it is not possible to scan a significant number of
models in the usual technique. Here we present a new way of
computing the CFs which is extremely efficient when many
models are needed. We first compute the statistics
of \emph{subhalo} pairs, and save them as a function of $\mfal$ and
$\mf$. Only then are the galaxy CFs computed. In this section we explain how the statistics of
subhaloes is defined and computed, the next subsection discusses the way this
information is used to model galaxies.

We want to constrain our models against the stellar mass function of galaxies. In order to do
so we will use the mass function of central and satellite subhaloes:
\begin{equation}
\phi_c(\mfal)=\frac{1}{V}\frac{\dd{N^c}}{\dd{\log\mfal}} \,,
\end{equation}
\begin{equation}
\phi_s(\mfal,\mf)=\frac{1}{V}\frac{\dd^2{N^s}}{\dd{\log\mfal} \, \dd{\log\mf}} \,.
\end{equation}
Here $V$ is the volume of the simulation box, and $N^c,N^s$ are the numbers of central and satellite
subhaloes respectively. In Fig.~\ref{fig:nsat_2d} we show the two dimensional mass function $\phi_s$,
for a model with $\adf=3$. A similar behaviour as presented here is valid for
$0.1\leq\adf\leq10$, where the low-mass contour can vary by $\sim0.5$ dex, and the contour
of massive subhaloes is hardly affected.

In order to compute the CFs of galaxies, we start by computing the number
of subhalo pairs, $N_p$, for the total sample of subhaloes. In case both subhaloes within the pair are central
subhaloes, we count this pair into $\psi_{cc}$:
\begin{equation}
\psi_{cc}(\mfala,\mfalb, r ) = \frac{1}{V^2} \frac{\dd^3{N_p^{cc}}}{\dd{\log\mfala}\, \dd{\log\mfalb}\,\dd{\log{r}}} \,.
\label{eq:psi_cc}
\end{equation}
Here $\mfala$, $\mfalb$ are the infall mass of the first and second
subhaloes in the pair, and $r$ is the distance between these subhaloes within
the $x$-$y$ plane. Distances are computed only within the $x$-$y$ plane
in order to compute the \emph{projected} auto-correlation function,
as is described below. In practice we divide the range in Log$\mfal$ and
Log($r$) into 100 and 50 bins respectively, and save $\psi_{cc}$ as a multi-dimensional
histogram.

In a similar way we define the pair statistics of central-satellite,
and satellite-satellite subhaloes,
\begin{equation}
\psi_{ss}(\mfala,\mfalb, \mfa, \mfb, r ) \,,
\end{equation}
\begin{equation}
\psi_{cs}(\mfala,\mfalb, \mfb, r ) \,.
\label{eq:psi_cs}
\end{equation}
Note that for satellite subhaloes the number of pairs
is saved as a function of both $\mfal$ and $\mf$.
This is done in order to properly model the dependence of
stellar mass on $\mf$. It should be emphasized that the statistics
of satellite subhaloes, i.e. $\phi_s$, $\psi_{ss}$, $\psi_{cs}$,
depend on the dynamical friction constant, $\adf$, and on the
location we adopt for unresolved subhaloes ($p$ and
$q$ from Eq.~\ref{eq:sat_loc}).


\subsection{Definition of a model - domains in stellar mass}
\label{sec:def_model}

Previous studies have often used an analytic functional form to describe the
relation between $\mfal$ and $\ms$. For example, \citet{Moster10}
suggested:
\begin{equation}
\ms = f \left( \mfal \right) =  c_1 \mfal \left[ c_2 \mfal^{c_3} +
 \mfal^{c_4} \right]^{-1} \,,
\end{equation}
where $c_i$ are all constants. A straight forward way
to extend this approach here would be to parameterize $f$ as a function of two
variables, $\ms=f(\mfal,\mf)$, and to use a different set of
parameters for modeling central and satellite galaxies. However, this
approach requires a priori knowledge of $f$, and the  resulting solutions
might be restricted by the specific functional form chosen.
This is especially true when $f$ is allowed to be different for
satellite and central galaxies, so the freedom in its functional
shape might be larger than what was found in previous studies (see the examples in
section \ref{sec:examples} below).

Here we suggest a new way to parameterize the relation between
$\mfal$ and $\ms$, which is motivated by the observational data.
The observed CFs of galaxies are computed over
mass bins of width 0.5 dex in stellar mass (hereafter
`domains'). For example, the CF for small mass galaxies is based
on galaxies with stellar mass in the range
\begin{equation}
\textrm{Domain 1: } \,\, [10^{9.27},\,\, 10^{9.77}] \,\, \msun\,.
\end{equation}
In order to model the CF in this range we only
need to know the values of $\mfal,\mf$ that correspond to
$\ms=10^{9.27},10^{9.77}$. These are defined by:
\begin{equation}
\label{eq:domain_bound}
f\left(\mfalo,\mfo \right) = 10^{9.27} \,, \,\,\, f\left(\mfalt,\mft\right) = 10^{9.77} \,.
\end{equation}
Such constraints imply that in case of a smooth and monotonic $f$, the
'boundaries' of the domain correspond to curves within the $\mfal,\mf$
plane. The relation that defines a boundary can thus be written as:
\begin{equation}
\label{eq:boundary}
\mfalo=U_1\left( \mfo \right) \,,\,\, \mfalt=U_2\left( \mft \right) \,,
\end{equation}
where $U_i$ is the curve function.
All subhaloes that are located between $U_1$ \& $U_2$ 
(their $\mfal,\mf$ masses follow the constraint $U_1(\mf)\!\leq\!\mfal\!\leq\!U_2(\mf)$)
should contribute galaxies to the CF of the first domain.

\begin{figure}
\centerline{ \hbox{ \epsfig{file=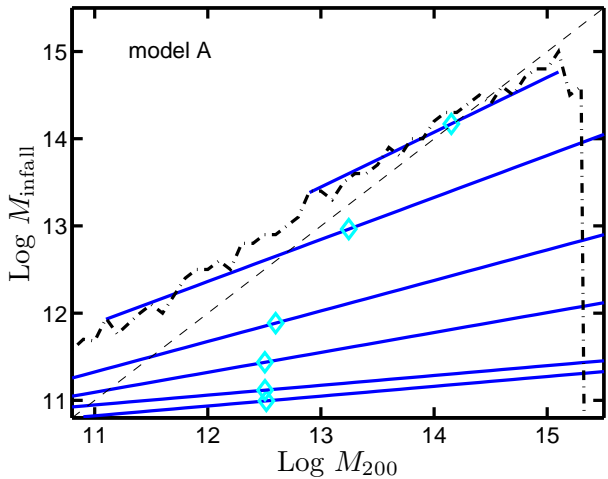,width=9cm} }}
\caption{
The dependence of $\ms$ on ($\mfal$,$\mf$) for satellite galaxies according to model $A$.
This model uses $\adf=3,p=0.5,q=0.8$.
The solid lines show $U_i^s$ that obey the equation $f\left(U_i^s(\mf),\mf\right)
=\msi$, where $\msi=$ 9.27, 9.77, 10.27, 10.77, 11.27,
11.77 in units of $\log\msun$.
Diamonds are placed at the
median $\mfal$ values along each line. The dotted-dashed line and the thin dashed
line are the same as in Fig.~\ref{fig:nsat_2d}.
}
\label{fig:model_a1}
\end{figure}

\begin{figure}
\centerline{ \hbox{ \epsfig{file=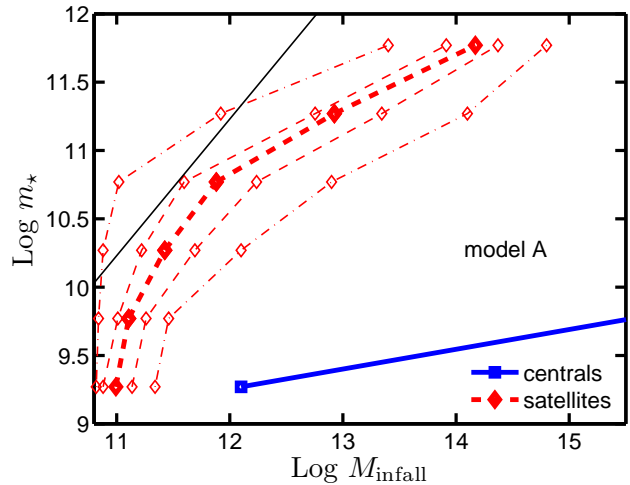,width=9cm} }}
\caption{
The relation between $\ms$ and $\mfal$ in model $A$.
For satellite subhaloes we plot various values of $\mfal$ per a given $\ms$, including
median (thick dashed lines), one standard deviation from the median (thin dashed lines), and
the full range (thin dotted-dashed lines).
The thin solid line corresponds to the universal baryonic fraction, $\ms=0.17\mfal$.
Model $A$ violates this fraction limit, and thus will not be used in the rest of the paper.}
  \label{fig:model_a2}
\end{figure}

The standard abundance matching approach assumes that $\ms$ depends
only on $\mfal$. In terms of our language, this means that the
functions $U_i$'s are all constants, with no dependence on $\mf$.
In this work, we extend this assumption in two ways: we assume that
$U_i$ might be a power-law for satellite galaxies\footnote{For central galaxies the values of
$\mf$ are always very similar to $\mfal$ so there is no added value in allowing the stellar mass
of central galaxies to depend on $\mf$.}, and that for each domain there
might be a different $U_i$ for central and satellite galaxies. This can be
summarized as:
\begin{eqnarray}
\label{eq:alpha_i}
\textrm{Central subhaloes:} & & U_i^c = \alpha_i \, ,\\ \nonumber
\textrm{Satellite subhaloes:} & & U_i^s = \beta_i \mf ^{\delta_i} \,.
\end{eqnarray}
Since the observed CFs are based on 5 domains in stellar mass, we
need to specify $U_i$'s at 6 domain boundaries. Our model therefore
includes 6 free parameters ($\alpha_i$) for central subhaloes,
and additional 12 parameters ($\beta_i,\delta_i$) for satellite
galaxies. We choose to limit our models only to power-law dependence
on $\mf$ following the results of paper I, and as a first order approximation. It will allow us to test
how important the standard assumption of constant $U_i$ is.
It might well be that a more complex behaviour would add a
significant amount of freedom to the models.

In order to model the stellar mass function, we
will use the same domains, and demand that the stellar
mass function will be reproduced once integrated over each domain. This guarantees
that in case it is needed, a detailed solution of the type $\ms=f(\mfal,\mf)$ exist.
However, since our models allow the mass relation above to deviate between central and
satellite galaxies, the detailed behaviour of $\ms$ within each
domain is not well constrained. In general, there might be many
different interpolations of the kind $\ms=f(\mfal,\mf)$ within each
domain (between adjacent $U_i$'s). These will not change the computed CF, and will fit the
observed stellar mass functions. The range in stellar
mass for each domain is relatively small, so this freedom is negligible in comparison to the
results we will show below.

In general, modeling a scatter in $\ms$ for a given $\mfal$ and $\mf$ is possible
within our formalism. However, it demands a detailed knowledge of the functional
form, $\ms=f(\mfal,\mf)$. This addition does not allow us to scan the different models
in a very efficient way. We therefore do not treat such a scatter in this work.
Nonetheless, the dependence of $\ms$ on $\mf$ for satellite galaxies
results in a variation in $\ms$ as a function of $\mfal$. This effect will
be discussed below.


\subsection{How to compute CFs?}
\label{sec:model_howto}

To summarize, each model in this work is defined by the following parameters:
\begin{description}
  \item[a)] The value of $\adf$ used in the dynamical
  friction formula, as defined by Eq.~\ref{eq:t_df}.
  \item[b)] The values $p,q$ used in Eq.~\ref{eq:sat_loc} for
  modeling the location of unresolved subhaloes. Alternatively, we use the
  location as given by the most bound particle.
  \item[c)] For central subhaloes, 6 values of $\alpha_i$ that
  define the domain boundaries $U_i^c$ (Eq.\ref{eq:alpha_i}).
  \item[d)] For satellite subhaloes, 6 values of $\beta_i$, and
  6 values of $\delta_i$ that correspond to the $U_i^s$ boundaries.
\end{description}

Once the parameters $\adf$, $p$, $q$ are chosen, we
construct the subhalo statistical functions $\phi$ and $\psi$. The
parameters in items (c) and (d) above are then used as integration limits
for $\phi$ and $\psi$.
By integrating $\phi$ we compute the total number of subhaloes within the domain:
\begin{eqnarray}
\label{eq:N}
\lefteqn{
N = V\int_{U_i^c}^{U_{i+1}^c} \phi_c\, \dd{\log\mfal} \, + } \\ \nonumber & &
V\int_{U_i^s}^{U_{i+1}^s} \phi_s \, \dd{\log\mfal} \, \dd{\log\mf}   \,.
\end{eqnarray}
A similar integration of $\psi$ over each domain results in $N_p(r_p)$ -- the total
number of pairs within each radial bin $r_p$.  The projected auto-correlation function,
$w_p(r_p)$, is then defined as the deviation
in the number of pairs from the average value per volume:
\begin{equation}
w_p(r_p) = \left[ \frac{L^2}{N^2} \frac{V^2\,N_p(r_p)}{A_p} - 1 \right] L \,.
\end{equation}
Here $A_p$ is the 2-dimensional area covered by the radial bin $r_p$, and $L$ is the size of
the simulation box in $h^{-1}$Mpc.

We have tested numerically that our
methodology agrees with the standard ABM approach for various different
models. In our approach the models are based on quantifying $\phi$ and $\psi$ only,
so we do not construct a full realization of galaxies per each model. This is in
contrast to standard ABM models, where a list of all subhaloes within a given simulation
is necessary.

\begin{figure}
\centerline{ \hbox{ \epsfig{file=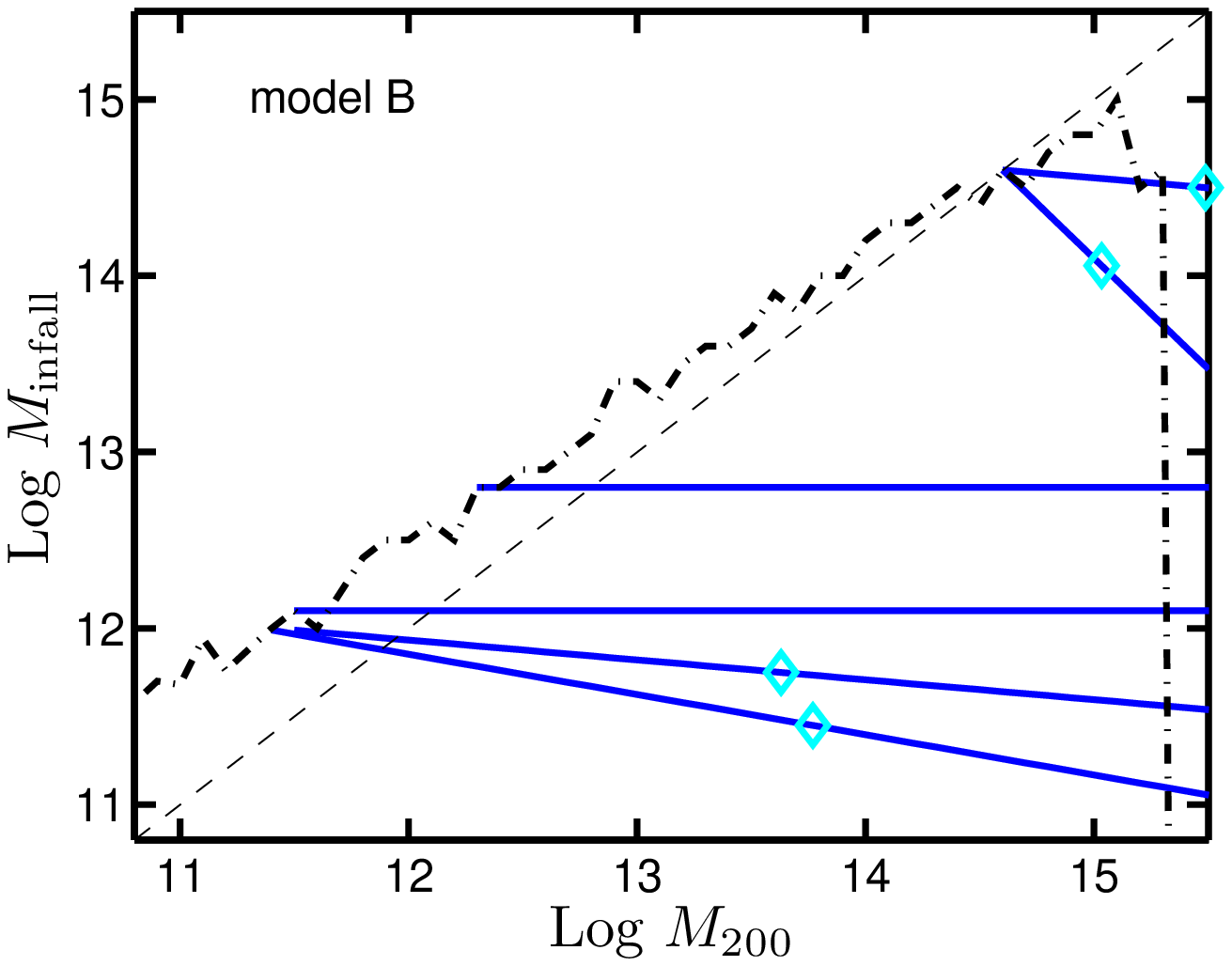,width=9cm} }}
\centerline{ \hbox{ \epsfig{file=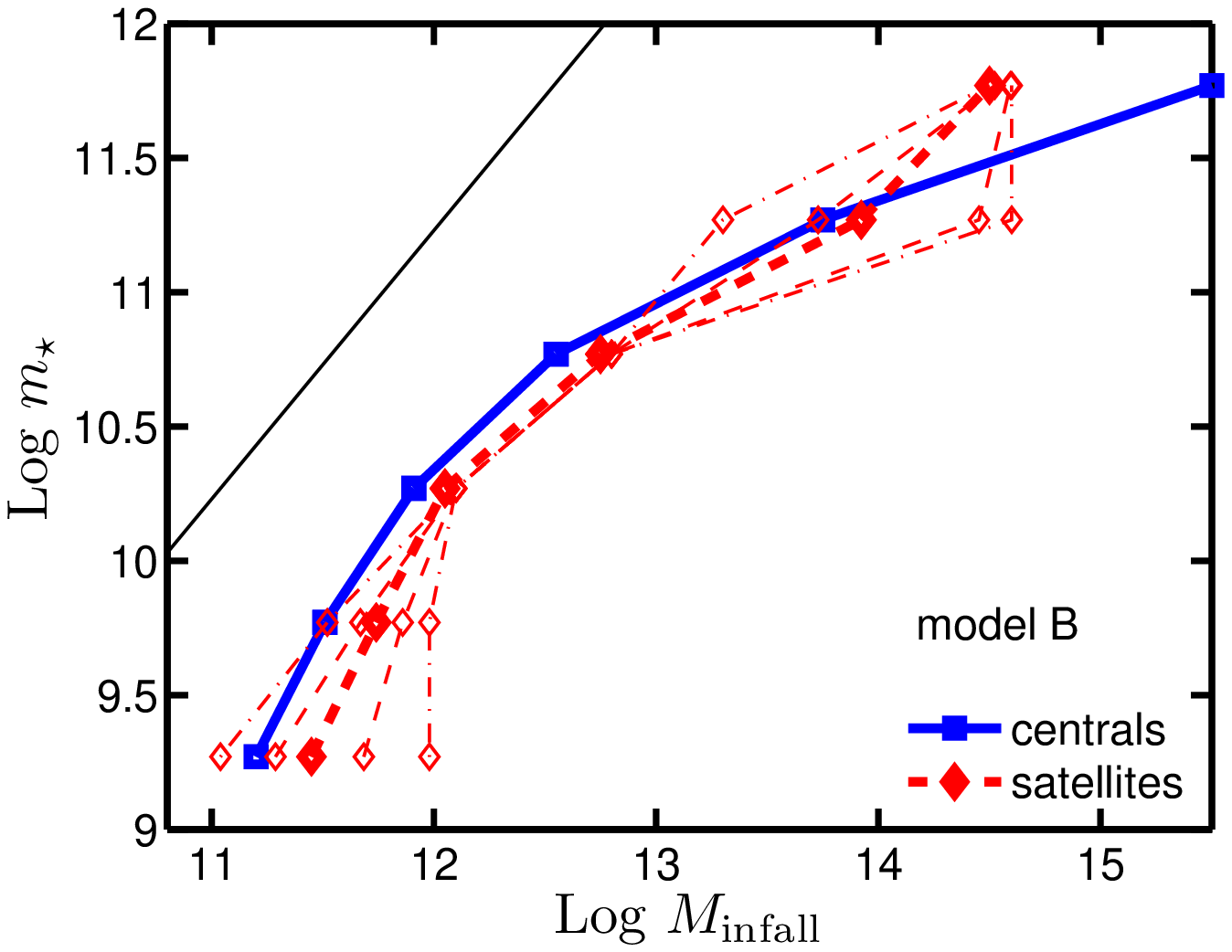,width=9cm} }}
\caption{Same as Figs.~\ref{fig:model_a1} \& \ref{fig:model_a2} but for model $B$. Here $\adf=3$,
and the location of unresolved subhaloes is set by the most bound particle.}
  \label{fig:model_b}
\end{figure}

\begin{figure}
\centerline{ \hbox{ \epsfig{file=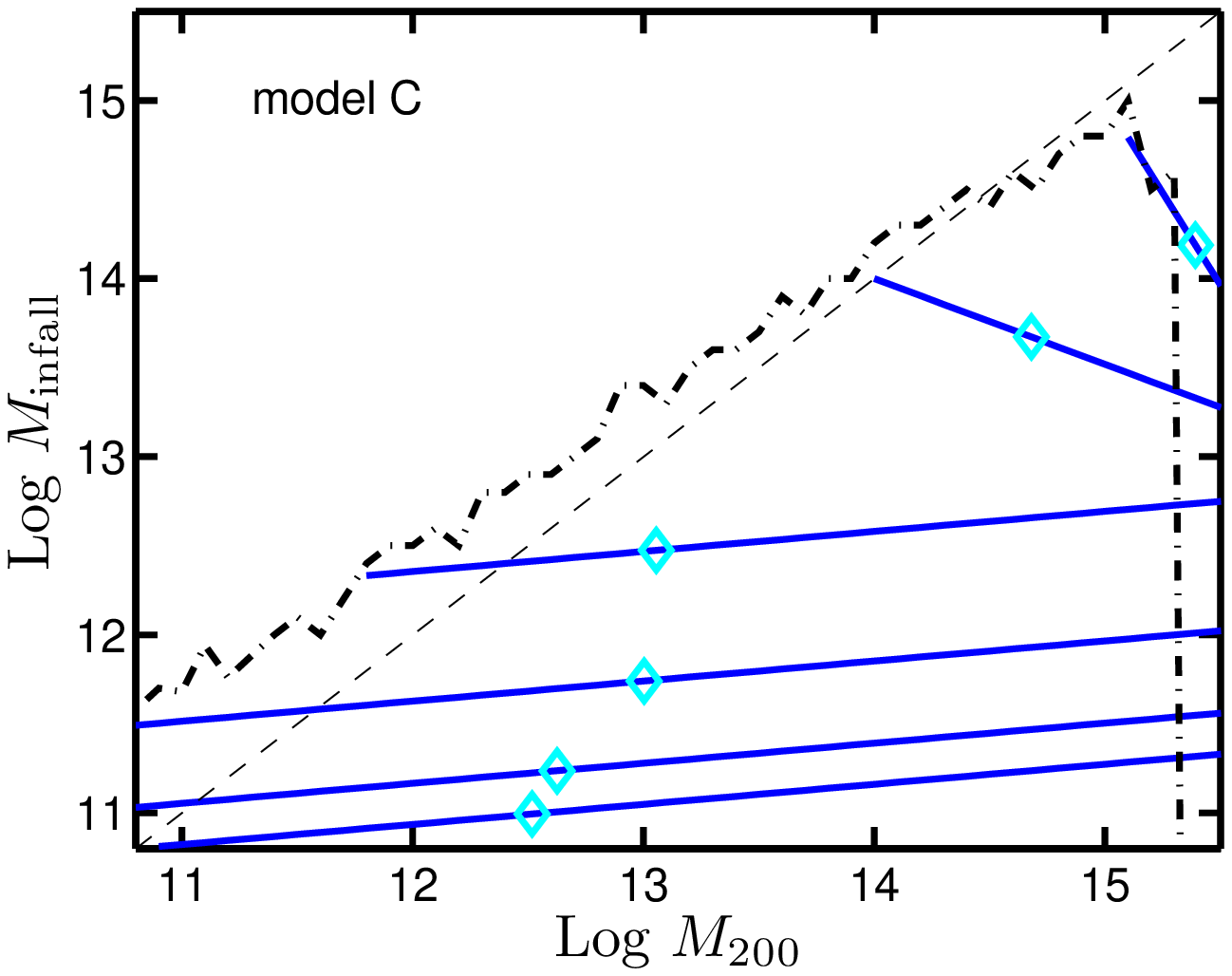,width=9cm} }}
\centerline{ \hbox{ \epsfig{file=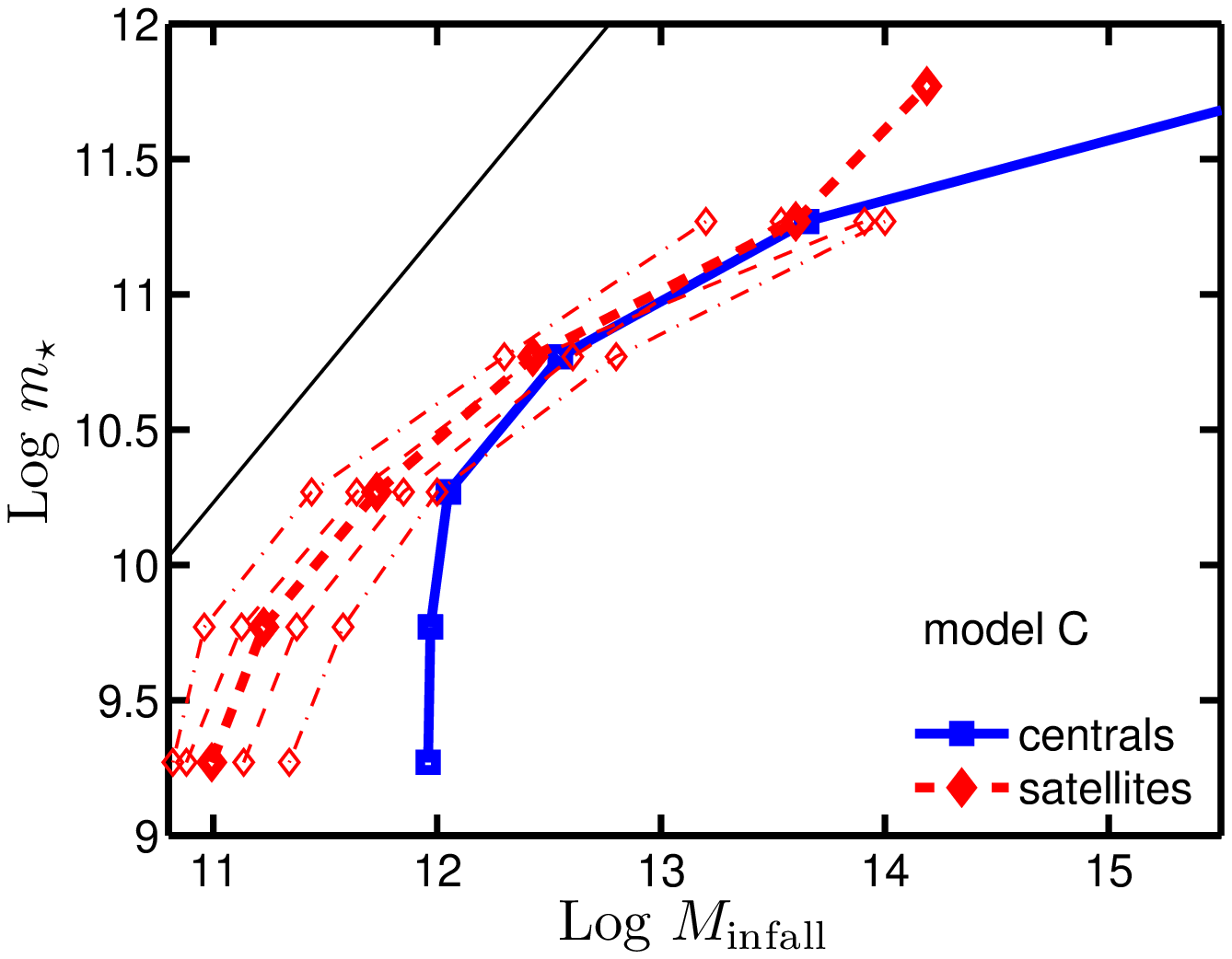,width=9cm} }}
\caption{Same as Figs.~\ref{fig:model_a1} \& \ref{fig:model_a2} but for model $C$. Here $\adf=3,p=0.5,q=0.8$.}
  \label{fig:model_c}
\end{figure}


\section{A few examples}
\label{sec:examples}

In this section we present a few specific models that were found using a large parameter search
within our formalism. The details of the search will be presented in the next section.
Here we discuss a few examples to demonstrate the parametrization needed for each model,
and to highlight the variety of models that are able to fit both the stellar
mass function and CFs.

In Figs.~\ref{fig:model_a1} \& \ref{fig:model_a2} we present model $A$, one example out of a family of models
which are extreme with respect to the full population of models.
In this model, $\ms$ is much higher for satellite subhaloes than for central subhaloes,
for a given $\mfal$. The difference can reach a factor of 100 in stellar mass.
As can be seen from Fig.~\ref{fig:model_a1}, this model includes a relatively strong dependence
of $\ms$ on $\mf$, a feature that was originally seen in the SAMs analyzed in paper I, although less
prominent. We note that previous ABM models were using only
horizontal lines in the $\mf-\mfal$ plane.

\begin{figure}
\centerline{ \hbox{ \epsfig{file=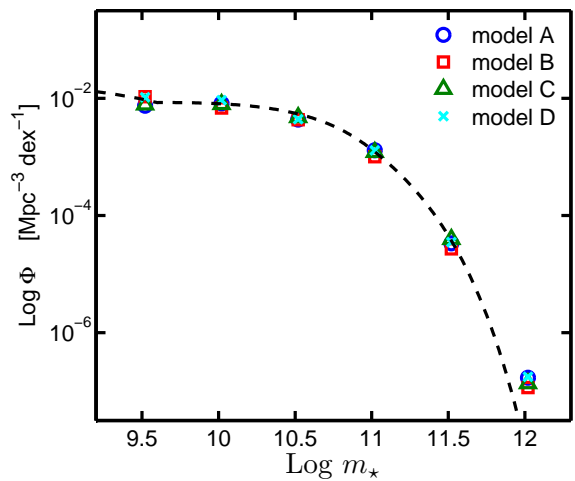,width=9cm} }}
\caption{
The stellar mass functions of galaxies using models $A,B,C,D$ (symbols).
The observed function derived by \citet{Li09} is plotted in dashed line.
Symbols are placed at the centre of each domain.
}
\label{fig:smf_examples}
\end{figure}

\begin{figure*}
\centerline{\psfig{file=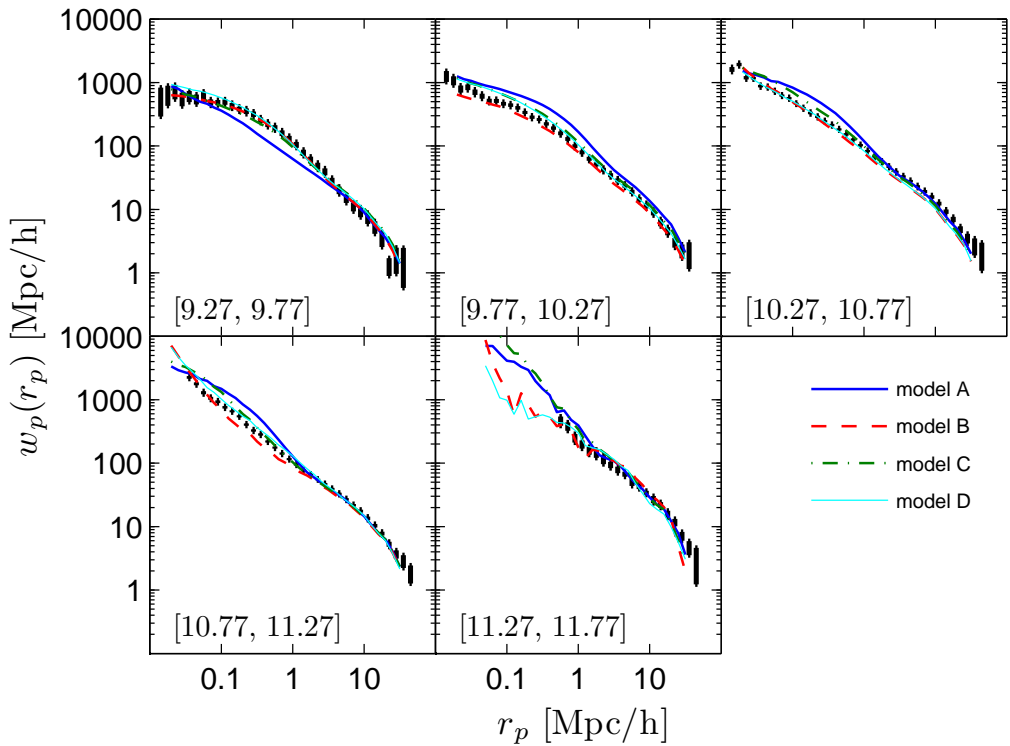,width=150mm,bbllx=30mm,bblly=80mm,bburx=188mm,bbury=200mm,clip=}}
\caption{The projected auto-correlation functions (CFs) derived for models $A,B,C,D$.
Each panel corresponds to galaxies with stellar masses as indicated by the range of Log$\msun$.
Lines show the results of the models. The observational
data are using SDSS DR7 with the same technique as in \citet{Li06}, and are shown as error bars.
Models $B,C,D$ fit the observational data to a level of 0.1 dex RMS, while model
$A$ deviates at the level of 0.2 dex (details on the way these errors are computed
can be found in section \ref{sec:strategy}).}
\label{fig:cfs}
\end{figure*}

The CFs of model $A$ are plotted in Fig.~\ref{fig:cfs}, showing a reasonable match to the
observed data, with an RMS deviation of 0.2 dex (more details on how we define this deviation
can be found in section \ref{sec:strategy}). Interestingly, the same model fits the stellar mass 
function well (see Fig.~\ref{fig:smf_examples}). These results suggest that even though model $A$ is extreme,
it is broadly consistent with the observational constraints adopted here.
However, Fig.~\ref{fig:model_a2} shows that this model violates the limit $\ms<0.17\mfal$
for satellite subhaloes. Assuming all baryons within each subhalo are converted into stars,
the maximum stellar mass should equal $\ms=\Omega_B \Omega_m^{-1} \mfal=0.17\mfal$. Model $A$ exceed
this limit, and is therefore rejected and will not be considered
as a valid model in what follows. It is plotted here to demonstrate
that using our formalism, fitting both the stellar mass function and the CFs is not enough
for constraining the maximum stellar mass per a given $\mfal$.

\begin{figure}
\centerline{ \hbox{ \epsfig{file=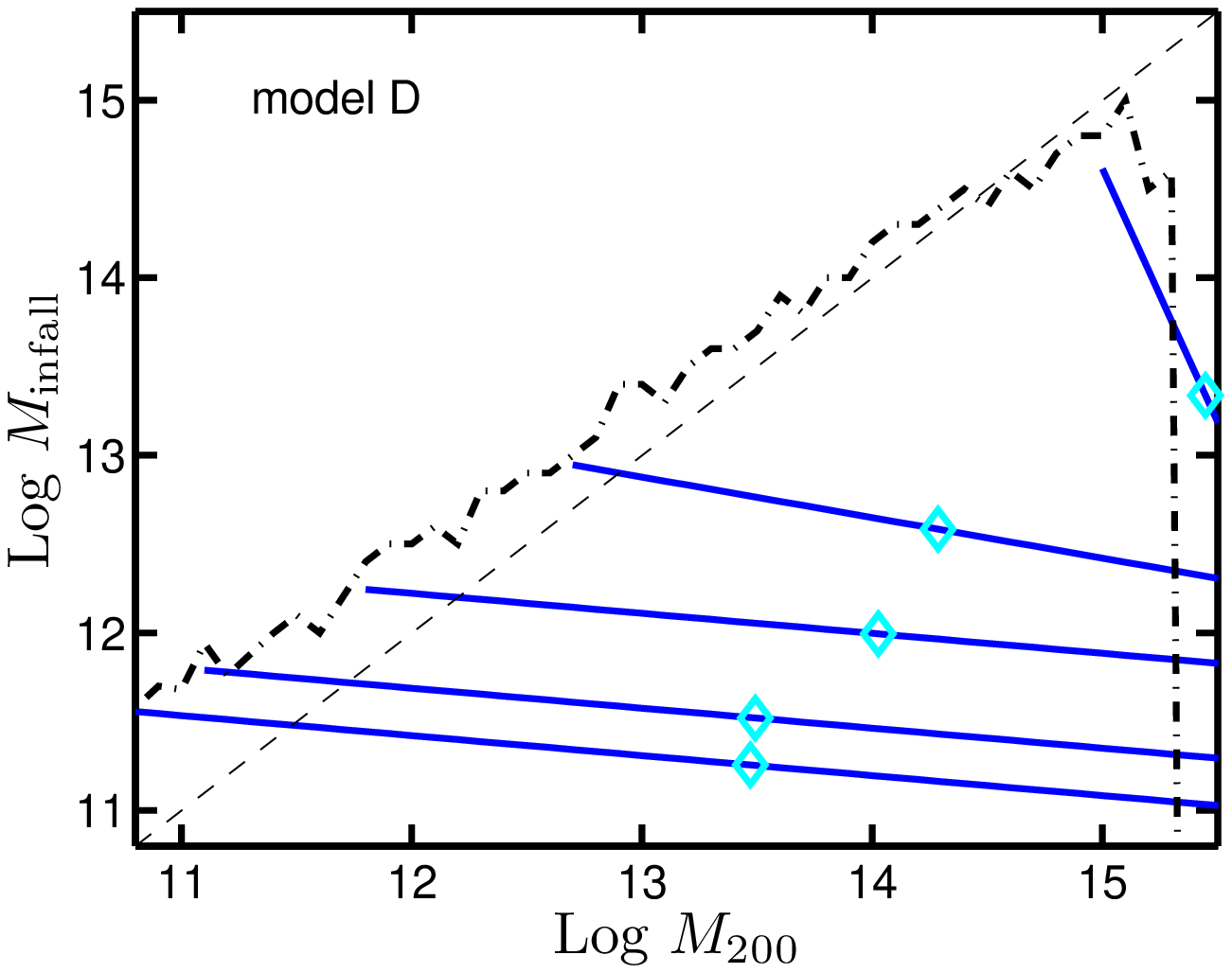,width=9cm} }}
\centerline{ \hbox{ \epsfig{file=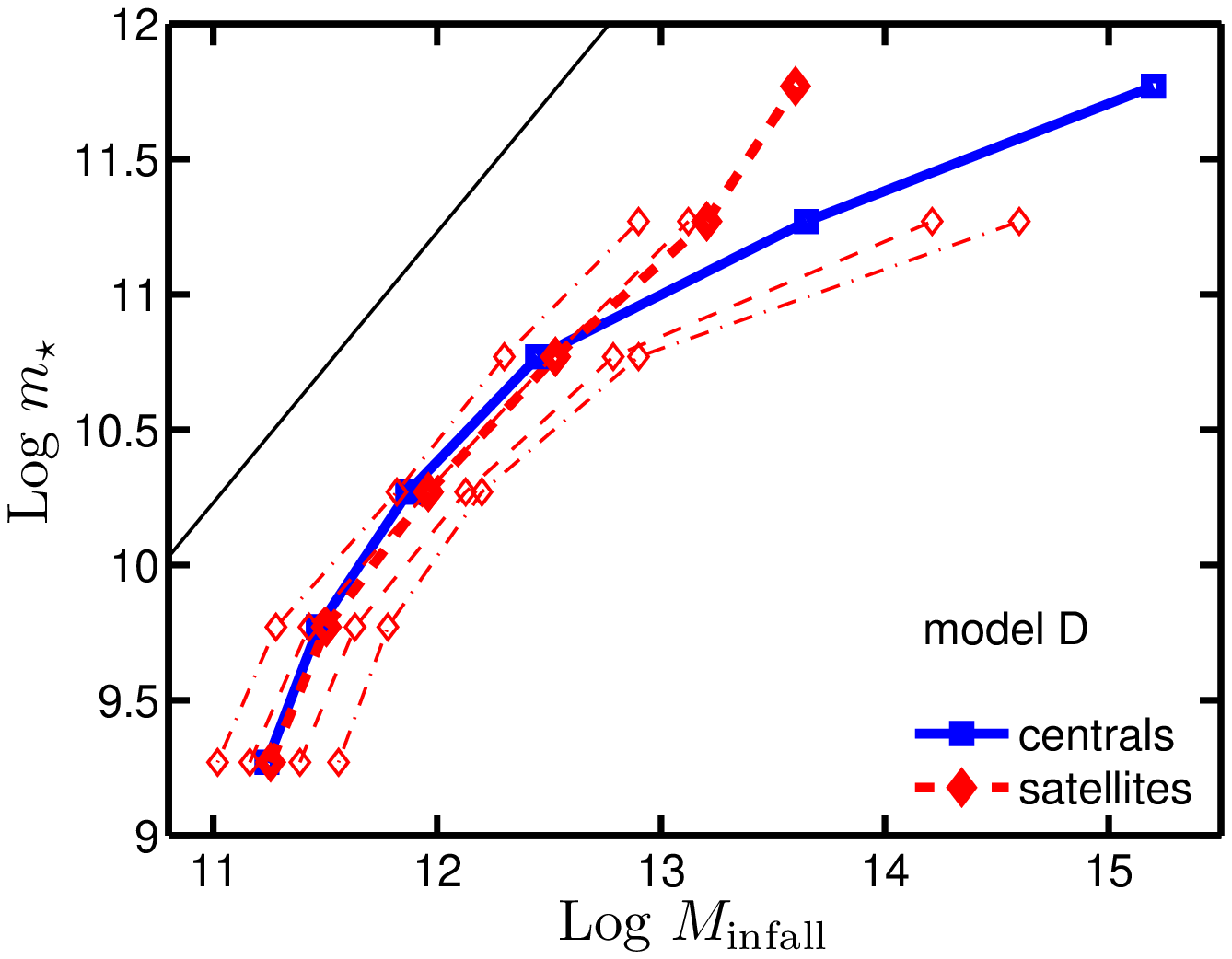,width=9cm} }}
\caption{Same as Figs.~\ref{fig:model_a1} \& \ref{fig:model_a2} but for model $D$. Here $\adf=3$,
and the location of unresolved subhaloes is set by the most bound particle.}
  \label{fig:model_d}
\end{figure}

\begin{figure}
\centerline{ \hbox{ \epsfig{file=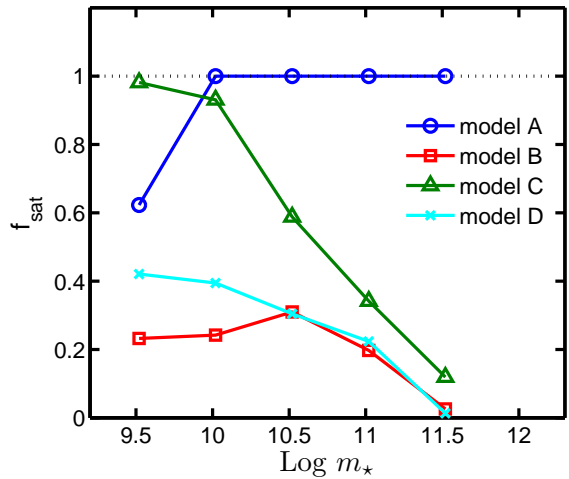,width=9cm} }}
\caption{
The fraction of satellite galaxies out of all galaxies, $f_{\rm sat} = \phi_s/(\phi_s+\phi_c)$,
for the different models presented here.
}
\label{fig:sat_frac}
\end{figure}

\begin{figure}
\centerline{ \hbox{ \epsfig{file=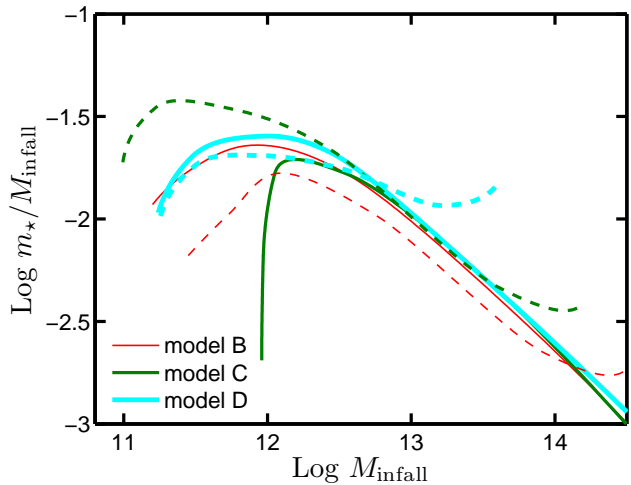,width=9cm} }}
\caption{The ratio between stellar mass and $\mfal$ for models $B,C,D$.
Satellite subhaloes are plotted as dashed lines, and are using the median values for a given $\ms$ as shown in
Figs.~\ref{fig:model_b}, \ref{fig:model_c}, and \ref{fig:model_d}.
Thin, medium, and thick lines correspond to models $B,C,D$ respectively. All lines are artificially smoothed
within the $\ms$ domains for a better view. The peak in the stellar fraction can change by more than a factor
of 10 between different models, and between satellite and central subhaloes.}
  \label{fig:mass_ratio_bcd}
\end{figure}

Models $B$ and $C$ are plotted in Figs.~\ref{fig:model_b} \&
\ref{fig:model_c}, and show
a very different behaviour for low mass subhaloes. The difference for a given $\ms$ between these two models
can reach a factor of 10 in $\mfal$ for central and satellite subhaloes. Model $C$ shows a
steep dependence of $\ms$ on $\mfal$ for low mass central subhaloes. This means that for
a given subhalo mass, the difference between $\ms$ for satellite and central galaxies might be
very large, more than a factor of 10.

We emphasize that our definition of a `satellite' versus `central' is valid only for subhaloes. It might
be that satellite subhaloes will host more massive galaxies than the central subhalo
within the same group. In these cases the more massive galaxy in the group
might be identified as a `central' galaxy, although its host subhalo is defined as a `satellite' here.
In a recent study, \citet{Skibba11}, have estimated the fraction of haloes that host central galaxies that
are not the most luminous in their group. Using a group catalog based on SDSS \citep{Yang07} they found that this fraction
reaches 25 (40) per cent for low (high) mass haloes. Thus, a different identification of `central' and `satellite' may be used either
in observational studies, or in the analysis of hydrodynamical simulations. In our models, the difference between satellite
and central subhaloes is defined solely according to the dark-matter behaviour, and is motivated by the
different merger history of subhaloes, and their final location.

From Fig.~\ref{fig:model_c} it is evident that model $C$ predicts a
non-negligible number of low mass subhaloes that host galaxies
with very high stellar masses, including galaxies with $\ms \sim0.6\,\cdot\,0.17\mfal$.
This shows again that our formalism cannot constrain the mass fraction that is locked in stars,
in contrast to what has been claimed using
more simplified ABM models \citep[e.g.][]{Behroozi10,Guo10a}.

Model $C$ is a good example of a model with a very high abundance of
satellite galaxies.
Although the number of satellite \emph{subhaloes} is smaller than central subhaloes at a given $\mfal$,
different $\msh$ relations for central and satellite subhaloes might result in
a high abundance of satellite \emph{galaxies}.
We plot the satellite fraction (the number of satellite galaxies out of all the galaxies of a given
$\ms$) of all the models in Fig.~\ref{fig:sat_frac}. The satellite
fraction changes significantly between the models.
For model $C$ this fraction reaches unity at low stellar masses, while for models $B$ and $D$ it reaches
a maximum value of $\sim0.3$ and 0.4 respectively.

Model $D$ is shown in Fig.~\ref{fig:model_d}. It is an example for a model in which
$\ms$ for satellites does not strongly depend on  $\mfal$.
In general, the slope of the $\msh$ relation, and the location
where it turns over can be related to various processes in galaxy formation
physics. Merger-rates, cooling, feedback and star-formation
should all be combined in order to reproduce the observed $\msh$
relation \citep[e.g.][]{Shankar06}. Unless our more extreme models like model $D$
are ruled out by other observational constraints,
our results indicate a large amount of freedom in modeling the above processes.

The stellar fractions, $\ms/\mfal$, for the few models presented here, are plotted
in Fig.~\ref{fig:mass_ratio_bcd}.
The halo mass at which the global stellar fraction reaches a maximum can range
from $\sim3\times10^{11}$ to $\sim3\times10^{12}\,\msun$. In model $D$ this efficiency
is approximately constant for satellite galaxies as a function of $\mfal$, showing no global
strong peak.

The CFs of all the models are shown in Fig.~\ref{fig:cfs}.
Models $B,C,D$ fit the observed CFs to a good accuracy, below 0.1 dex RMS,
while model $A$ is less accurate, reaching a level of 0.2 dex RMS. This demonstrates the importance
of using this constraint here, in order to narrow down the range of accepted
models. As can be seen from
Fig.~\ref{fig:smf_examples}, all the models fit the stellar mass
function well.
Another point that should be emphasized regarding Fig.~\ref{fig:cfs}
is the small scale ($<1$Mpc) clustering of massive
galaxies. The CF of different models shows variations at this regime, so
it might be that extending the CF of the most massive domain into
smaller radii will help to constrain the models. This will require
a survey volume much larger than in the current SDSS survey.


\section{Results}
\label{sec:results}

\subsection{Search strategy}
\label{sec:strategy}

Each model within our formalism is defined by 21 free parameters,
where 18 of them define the domains in stellar masses ($U_i^c$ \& $U_i^s$ from Eq.~\ref{eq:alpha_i}) used to compute
the CF, 1 parameter fixes the time-scale for dynamical friction, and 2 parameters
are responsible for the location of unresolved subhaloes. We were able to scan
\emph{all} possible options of the 18 parameters that govern the domains behaviour
(within the resolution adopted here). In addition, we have tested five
$\adf$ values in the range 0.1--10,
and a few options for setting the location
of unresolved subhaloes (see Appendix \ref{sec:sat_loc} and Table \ref{tab:models} for more details).
In total, we have tested the stellar mass function for
$\sim10^{12}$ models, approximately $10^7$ out of them were tested
against the observed CFs.
More numerical details regarding the search algorithm can be found in
Appendix \ref{sec:scan_prm}.

The values of $\phi$ and $\psi$ are saved numerically in fine bins. The choice of bin size is important
for several reasons. First, the function $\psi_{ss}$ depends on 5 different variables, so the number of bins
we use for each variable is limited because of computer memory issues. Second, fine bins will require
more evaluations of CFs, as the code checks automatically all possible options. This can slow our code
dramatically. On the other hand, the bin size
limits the search resolution: too big bins will not allow us to find all possible solutions.
As a compromise between these requirements we have chosen the following bin sizes:
$\mfal$ is split into bins of 0.02 (0.1) dex for subhaloes with mass smaller (bigger) than
$10^{12}\,\msun$, data per $r$ and $\mf$ are saved in bins of 0.1 dex.
We use small bins for low mass $\mfal$ because the typical difference between adjacent $U_i$ is smaller
for small mass $\mfal$ (see e.g., model $C$ above).

As described in section \ref{sec:def_model}, the domain boundaries 
are defined using power-law relations. Writing Eq~.\ref{eq:alpha_i} in terms of Log mass gives
$\log\mfal=\log\beta_i+\delta_i\log\mf$. Our search
algorithm checks all the possible values of $\beta_i$ according to the bins of $\mfal$.
$\delta_i$ is modeled in terms of the line slope, $\delta_i=\tan \theta_i$. We sample
$\theta_i$ in steps of $6.75$ degrees, over the range $[-90,\, 45]$.

Each model is accepted if it follows the conditions below:
\begin{itemize}
 \item The model fits the stellar mass function of \citet{Li09} with an accuracy that is better than 20 per cent.
This criterion is applied separately to each domain by integrating the stellar mass function over
the domain range (0.5 dex in $\ms$). We have checked that the range of models
shown below changes in a very minor way, when demanding a better fit that resembles the statistical
errors from \citet{Li09} (these are 5,5,5,10,20 per cent for each domain, in order of increasing mass).
Here we choose to use a constant accuracy of 20 per cent to account for systematic uncertainties in the 
stellar mass function \citep[comparisons against other measurement can be found in][]{Guo10a,Bernardi10}.
\item The model fits the logarithm of the observed CF (see Fig.~\ref{fig:cfs}) to better than an RMS value of 0.1 dex
(26 per cent). This estimate is based on all points in the range $0.03<r<30$ Mpc $h^{-1}$, sampled in bins separated by
0.2 dex in Log($r$). In order to test the effect of this fit accuracy we also show results using 0.2 dex deviation.
A detailed discussion of these issues is given below.
\item The model does not include individual points that deviate from the observed CF to more than a factor of 2.
\item The stellar mass function is fitted also
for masses larger than the most massive domain (i.e. for~$\ms>10^{11.77}\,\msun$).
This domain should include only 23 galaxies, when using the full Millennium simulation.
Since this is a very small number, we allow
our models to deviate an additional Poisson error from the nominal value. This means that
the number of galaxies with masses larger than  $10^{11.77}\,\msun$ can have any value
in the range [14, 34].
\item The stellar fraction (i.e. $\ms/\mfal$) does not exceed the universal fraction of
0.17 for all subhaloes in the sample.
\end{itemize}

We adopt an accuracy of 0.1 and 0.2 dex for fitting the CFs due to the following reasons.
First, the Millennium simulation being used here is based on a cosmological model with
$\sigma_8=0.9$, higher than the most updated measurements of $\sigma_8=0.8$ \citep{Jarosik11}.
This should give rise to some deviation between our models and observations. We find that
there is a minimum deviation of 0.08 dex between all the models discussed here,
and the observed CFs. Since this deviation is dominated by scales larger than 1 Mpc, it is probably
related to the different cosmological model assumed. 
Second, due to the finite bins that are used for saving $\psi$, the accuracy of our search algorithm 
is approximately 0.03 dex. As a result of the above, our minimum fit criterion is chosen to be 0.1 dex.
However, since the error bars presented for the observed CFs are only statistical, it might be that systematic
uncertainties would introduce errors that are larger than 0.1 dex. 
The CFs depend strongly on various galaxy properties like color, star-formation, and morphology \citep{Li06}.
Consequently, if the stellar masses of galaxies
are systematically biased for galaxies of a given property (for example the high star-forming
galaxies), this could change the CFs considerably. 
We therefore additionally consider a fitting criterion of 0.2 dex, to demonstrate the effect of the
possible systematic uncertainties.

The minimum subhalo mass ($\sim6.3\times10^{10}\,\msun$),
enforced by the Millennium simulation seems to limit our models (see models $A$ and $C$ above).
It might be that using simulations of higher resolution will permit more models.
The Millennium II simulation \citep{Boylan09} is a good candidate for such a study. However,
high resolution simulations are naturally based on much smaller volume than what is being used here,
resulting in a smaller statistical sample and non-negligible cosmic variance effects. Since our study is
aiming at fitting the CFs to a high accuracy of 0.1 dex, we focus our study only on the Millennium simulation.


\subsection{The mass relation}

\begin{figure}
\centerline{ \hbox{ \epsfig{file=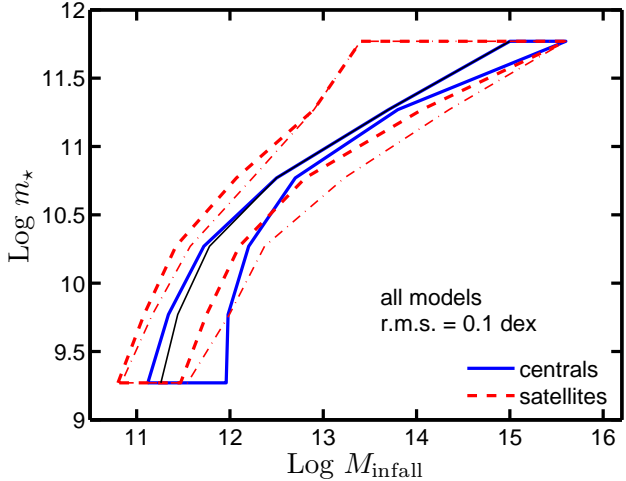,width=9cm} }}
\caption{A summary of all the models that fit the observed stellar mass function and the auto correlation functions
of galaxies at $\z=0$. Lines show the envelope of all relations between $\ms$ and $\mfal$ (we use \emph{median}
values of $\mfal$ per a given $\ms$ for each model). The results are separated into central and satellite subhaloes.
Thin dotted-dashed line corresponds to the median $\mfal$ for all satellite galaxies residing in haloes more
massive than $10^{15}\,\msun$. The models
reproduce the observed stellar mass function to better than 20 per cent, and the CFs to better than 0.1 dex RMS. The thin solid
line shows the $\msh$ relation using the same behaviour for satellite and central galaxies, with no dependence on $\mf$.}
  \label{fig:models_all}
\end{figure}

\begin{figure}
\centerline{ \hbox{ \epsfig{file=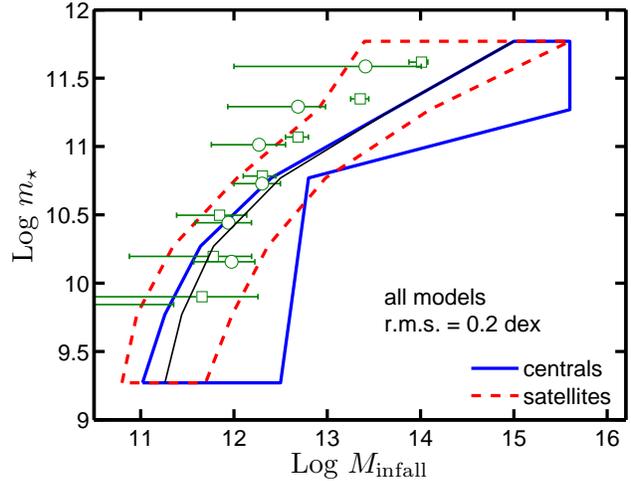,width=9cm} }}
\caption{Similar to Fig.~\ref{fig:models_all} but allowing the models to deviate from the
observed CFs by up to 0.2 dex RMS.
Results from weak lensing analysis by \citet{Mandelbaum06} are plotted in symbols. Squares and circles refer to the
\emph{mean} value of early and late type galaxies respectively, with error bars that reflect 95 per cent confidence level.
The fraction of late-type galaxies out of the full sample is 0.74, 0.60, 0.46, 0.32, 0.20, 0.11, 0.05
(ordered in increasing $\ms$). }
  \label{fig:models_58}
\end{figure}

The main results of the parameter search are summarized in Figs.~\ref{fig:models_all} and
\ref{fig:models_58}. In case we force the models to reproduce the observed CFs to a high
accuracy of 0.1 dex, the range of accepted models occupies a region of  $\sim$1 dex in the
$\msh$ plane, as shown in Fig.~\ref{fig:models_all}.
 Interestingly, the uncertainty is small for central massive subhaloes. On the other
hand, low mass central subhaloes can host galaxies with very low stellar masses. The range of
accepted models increases significantly in Fig.~\ref{fig:models_58}, where the accuracy of fitting
the CFs is set to 0.2 dex. We have checked that the distributions of models within
the plotted envelope in Figs.~\ref{fig:models_all}-\ref{fig:models_58}
are roughly uniform, so the range of models is not affected by outliers.

The number of models presented in Fig.~\ref{fig:models_all} and \ref{fig:models_58} is quite large.
Each domain includes $\sim10^5$ successful models in Fig.~\ref{fig:models_all}, and $\sim10^6$ models in
Fig.~\ref{fig:models_58}. We only take into account domains 
for which nearby $U_i$ from adjacent domains coincide, and the set of
all 5 domains covers the full mass range. Due to the above, we compute the median $\mfal$
for a given $\ms$, which does not demand combinations of domains of different $\ms$. 
Computing the opposite relation, median $\ms$ per a given $\mfal$, is much more complicated 
within our formalism. However, since the relation for central galaxies
does not include any scatter, it represents a median $\ms$ per a given $\mfal$ as well.

In our models, the stellar mass of satellite galaxies depends on their host
halo mass, which correlates with the
number of neighboring galaxies in observational studies. As a result, the range in the $\msh$ relation changes
as a function of the host halo mass. This is shown in Fig.~\ref{fig:models_all} by plotting the envelope of all median
relation in the $\msh$ plane, but taking into account only galaxies that reside in clusters (their host halo mass
is bigger than $10^{15}\,\msun$). For these satellite galaxies, the range in $\mfal$ for a given $\ms$ is larger
than the one for all galaxies.

Fig.~\ref{fig:models_58} includes the results of weak lensing analysis
from \citet{Mandelbaum06}.
These estimates are for the host halo mass\footnote{\citet{Mandelbaum06} define the halo mass as the mass
enclosed within a radius that corresponds to 180 times the mean density, this mass is higher than $\mf$ \citep[see also][]{Weinmann06a},
but agrees quite well ($<0.1$ dex) with $\mfal$ for central subhaloes used here.} of central galaxies, and should be compared to the solid lines
of our models. The square symbols represent a subsample of early type galaxies, while the circles correspond
to late-type galaxies. As is mentioned in the figure caption, the fraction of late-type galaxies is
very small at the most massive stellar mass bin (0.05).

Surprisingly, our results deviate from those of \citet{Mandelbaum06} at the high mass end, even when
the range of models is large, using an uncertainty of 0.2 dex in matching the CF.
This is similar to what was found by previous ABM studies, summarized by
\citet{Behroozi10}. Apparently, for a given value of $\ms$,
the weak lensing results constrain the host halo mass by providing mainly an upper
threshold (the low values of the host halo often reach the universal baryonic fraction).
The opposite seems to be true for central galaxies in our analysis.
Deviations between the two studies can be due to various effects:
\begin{itemize}
 \item \citet{Mandelbaum06} use a specific halo model for computing the lensing
signal, which differs from the set of models being used here.
\item There is some uncertainty
in their study owing to the width of the $\mfal$-distribution, for a given $\ms$.
\item Their stellar mass estimates are based on \citet{Kauffmann03} while the observations
of \citet{Li09} use the method of \citet{Blanton07}.
The difference between these estimates is discussed in both \citet{Li09} and \citet{Guo10a}, and its effect
on the stellar mass function is probably small. However, it might be that the difference between the methods
are more significant for computing the CFs.
\item It might be that our range of models is too narrow, or
that there are some systematic differences between the two approaches.
This could be due to our search resolution, the assumed
underlying cosmological models, or the assumed IMF.
\end{itemize}
A better, more self-consistent way to compare our results against weak lensing would demand
a direct estimate of the lensing signal from our models. This can then be compared to
the observed shear signal. Such an analysis is however outside the scope of this work.
A recent study by \citet{Leauthaud11} have used a model based on halo occupation distribution
and demonstrated the strength of applying this additional constraint.

The results of \citet{Mandelbaum06} indicate that the population of late-type massive central galaxies live inside
haloes with  lower masses than early-type galaxies.  The difference between early and late
type galaxies indicates that the scatter in the $\msh$ relation might be significant.
A large scatter was also found by \citet{More11}, based on satellite kinematics. We do not model such a scatter
for central galaxies in this study. Previous works \citep[e.g.][]{Moster10} have shown that including a scatter modifies
the $\msh$ relation at the massive end, such that for a given $\mfal$, $\ms$ decrease with increasing scatter.
Interestingly, the results of \citet{Mandelbaum06} reach the universal
baryonic fraction, $\ms/\mfal\sim0.17$, for part of the galaxy population. This is very similar to what
is found here, and can be seen in model $C$ (Fig.~\ref{fig:model_c}).

\begin{figure}
\centerline{ \hbox{ \epsfig{file=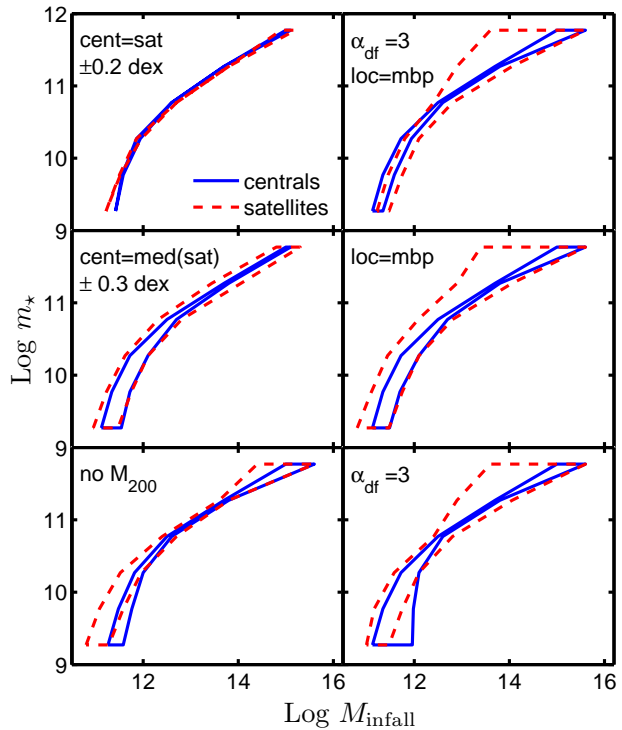,width=9cm} }}
\caption{The effect of different parameters on the median relation between $\ms$
and $\mfal$. Each panel shows a selection of models out of the full sample, as
plotted in Fig.~\ref{fig:models_all}.
The \emph{upper left} panel takes into account all models for which the behaviour
for satellite and central subhaloes is the same ($\pm0.2$ dex), and no dependence on $\mf$ is allowed.
In the \emph{middle left} panel we plot models with a similar median $\mfal$ for
central and satellite galaxies, at a given $\ms$. The \emph{lower left} panel plots models
for which there is no dependence on $\mf$.
Models using the default value of $\adf=3$ and location set by the most bound particle
are plotted in the \emph{upper right} panel. The \emph{middle right} panel summarizes models
that use the location as given by the most bound particle.
The \emph{lower right} panel summarizes all models with the same dynamical
friction constant, $\adf=3$.}
  \label{fig:models_cuts}
\end{figure}

In Fig.~\ref{fig:models_cuts} we plot various sub-samples of the models from Fig.~\ref{fig:models_all}, showing the influence
of different parameters on the relation between $\ms$ and $\mfal$. Our results agree with previous studies in predicting
a very tight $\msh$ relation, once satellite and central galaxies are considered to have very similar
median relations (at the level of $\sim0.2$ dex), and $\mf$ does
not affect $\ms$. This tight relation holds even though the models include the new ingredients related to $\adf$ and
the location of satellite subhaloes. This proves that
the dependence of $\ms$ on $\mf$, and the difference between central and satellite galaxies,
are responsible for the range of allowed models shown in Figs.~\ref{fig:models_all}-\ref{fig:models_58}.

It is worth noting that our models do not include solutions that have \emph{exactly} the same
mass relation for central and satellite galaxies, and can still fit the CFs to a level of 0.1 dex. The minimum deviation
between central and satellite galaxies in the upper left panel of Fig.~\ref{fig:models_cuts} is roughly 0.12 dex in $\mfal$.
This small difference might be an artifact of the
underlying cosmological parameters assumed by the Millennium simulation.

The models shown here use the two populations of galaxies to compensate for each other, keeping the overall $\msh$ relation
relatively similar. This effect is shown in the middle left panel of Fig.~\ref{fig:models_cuts}. Once we force the models
to have a similar median $\mfal$ for satellite and central galaxies, the range of models decreases, especially at the high
mass end. It might be that constraints on the differences between $\ms$ of the two populations, for a given $\mfal$, could be
enforced ad-hoc. However, it is not clear what this limit should be. A reasonable demand would be that the full distribution
of the two populations would overlap to some level, allowing a transition of central galaxies into satellites. This condition
seems to be fulfilled by our models (see Figs.~\ref{fig:model_b}-\ref{fig:model_d}). Since we do not model a distribution
for the $\ms$ of central galaxies, we do not explore this issue further here.

The effect of $\adf$ and satellite locations is explored in Fig.~\ref{fig:models_cuts}. It seems that both elements add freedom
to our models, although the effect of satellite location is slightly stronger. Higher resolution cosmological simulations
might help to constrain $\adf,p,q$. However, these simulations are based on dark-matter only, while the results here
are affected strongly by the baryonic components of galaxies \citep[see e.g.][]{Boylan08,Jiang08}.

\begin{figure}
\centerline{ \hbox{ \epsfig{file=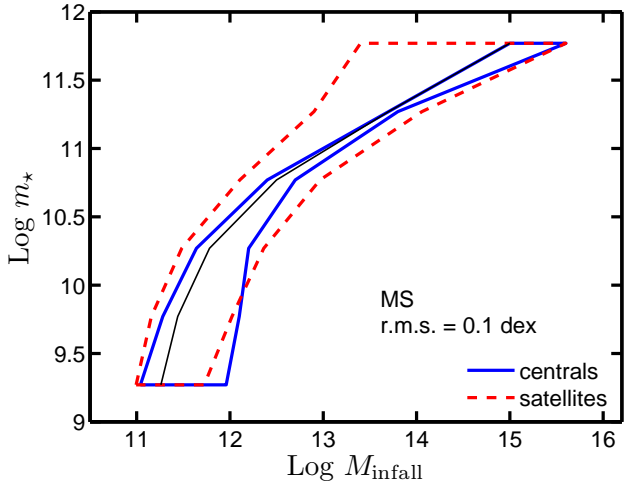,width=9cm} }}
\caption{Similar to Fig.~\ref{fig:models_all}, but here the CFs are constrained to fit
one of our models, instead of the observational
data. We use as a reference the simplest model where satellite and central
galaxies have the same relation between $\ms$ and $\mfal$, and $\adf=3$.
This shows that the uncertainty in the median relation between $\ms$ and $\mfal$
does not depend strongly on the reference model, and its underlying cosmological parameters.}
  \label{fig:models_ms}
\end{figure}

Lastly, in Fig.~\ref{fig:models_ms}, we show the results of a
parameter search when the reference CFs are not the observed ones,
but instead are taken from one model out of the models tested here. The range of accepted models
in this case is related to the internal degrees of freedom of the models, and is less related to the choice
of the cosmological parameters, the method adopted for computing $\ms$, or to the specific physical assumptions used in our formalism.
The fact that Fig.~\ref{fig:models_ms} is similar to Fig.~\ref{fig:models_all}
indicates that the underlying cosmological model does not affect the
range of acceptable models significantly. 
This is because here the reference CFs are based on the same cosmology, but use a specific 
model for the relation between $\ms$ and subhaloes.
As was pointed out by \citet{Cacciato09}, the effect of the underlying cosmological model should
be important when constraints from abundance and clustering of galaxies are combined with 
galaxy-galaxy lensing measurements. This issue should be further examined using
an appropriate $N$-body cosmological simulation.

\section{Summary and discussion}
\label{sec:discuss}

In this work we have studied the relation between the stellar mass of
galaxies ($\ms$) and the mass of their host subhaloes ($\mfal$). Our models are
constrained by the abundance of galaxies, and their auto-correlation function (CF) at
$\z=0$. We have shown that
once the population of galaxies is broken into
two sub-populations of central and satellite galaxies,
the allowed range in the $\msh$ relation
for each population reaches a factor of $\sim10$. The range of accepted models
depends on the accuracy by which the models reproduce
the observed abundance and clustering of galaxies.
As was demonstrated in our previous work
\citep[][paper I]{Neistein11}, a different $\msh$ relation for central and
satellite galaxies is expected in galaxy formation models, although
the strength of the deviation cannot be constrained easily.

The shape of the $\msh$ relation has a large degree of freedom, resulting in a
stellar fraction  (the ratio between
$\ms$ and $\mfal$) that can peak anywhere between $\sim3\times10^{11}$ and $3\times10^{12}\,\msun$. It
can even be close to
a constant as a function of $\mfal$ for satellite
galaxies. Interestingly, the stellar fraction can reach high
values, comparable to the maximum universal value of 0.17, for low mass satellite galaxies
residing in low mass haloes.
Although it seems unlikely that the stellar fraction would exceed 0.17, our models
are not able to constrain this ratio, and instead we are forced to use it as an ad-hoc constraint.

Our models are similar to the `abundance matching' (ABM) approach, and are directly relating a
galaxy with a specific $\ms$ to each subhalo within a large $N$-body simulation.
However, in comparison to previous studies, we include various additional ingredients for modeling
satellite galaxies. Their stellar mass
depends on both the host subhalo mass ($\mfal$) and on the halo
mass at $\z=0$, and might differ from the stellar
mass of central galaxies of the same $\mfal$. We include subhaloes
that have merged into more massive subhaloes at high-$\z$ but might host
galaxies that will survive until $\z=0$, due to long dynamical friction time-scales.
The location of these `unresolved' subhaloes is set by either the
location of their most bound particle, or by an analytical model.
We found that the most important ingredient for fixing the range of accepted models
is the way stellar masses for satellite galaxies are modeled.

A model with two populations of galaxies includes a large amount of freedom in
matching the observed stellar mass function. We therefore constrain
our models also against the observed auto-correlation function (CF) of
galaxies. We developed a new formalism to compute the model CF based
on the pair statistics of subhaloes. This enabled us to scan
systematically a significant part of the parameter space. We have
tested $\sim10^{12}$ different ABM models, and estimated the CF for
$\sim10{^7}$ models that showed a good match to the stellar mass
function. The accuracy by which we fit the CF affects
significantly the range of accepted models within the $\msh$ plane.
Consequently, a detailed study of the systematic uncertainties involved in
this measurement is crucial in order to better constrain our models.

The range allowed in the $\msh$ relation might be larger than our
prediction due to various effects. There are still a large number of
models that we did not test here. These include complicated
dependence of $\ms$ on the halo mass, different
estimates for dynamical friction and satellite locations,
higher resolution of the underlying $N$-body simulation, different
observed stellar mass functions \citep[e.g.][]{Bernardi10},
and modeling a random scatter in $\ms$. On the other hand, the range of
models might be modified once we use subhaloes from an $N$-body simulation using
the most up-to-date cosmological parameters.
Previous estimates for the uncertainty in the $\msh$ relation were emphasizing
the contribution from uncertainty in
the stellar mass function, resulting in 0.25 dex
\citep{Behroozi10}. This uncertainty should be added to what we find here.

In this paper we have reproduced the abundance
and clustering of galaxies only at $\z=0$ when constraining the $\msh$ relation. It is not likely that data from higher redshifts
would limit the $\msh$ relation to a narrow range.
This is because the same uncertainties discussed here would be valid at high-$\z$, in
addition to the larger observational errors inherent at these redshifts.
Moreover, once the $\msh$ relation is fixed at some redshift range,
it is not straightforward to decide which models violate the physics of galaxy formation
by linking galaxies at different epochs. For example, as we showed
in paper I, the $\msh$ relation for satellite galaxies at the time
of infall might already be different from the relation for central
galaxies at the same epoch.

Our empirical results have implications for various aspects of galaxy formation.
\citet{Guo10a} and \citet{Sawala11} argued that stellar fractions at a fixed subhalo
mass derived from basic ABM models are systematically lower than the
results of detailed hydro-simulations. Our results that allow for a much larger spread
of stellar mass at a fixed subhalo mass, are less definitive in this respect. This
is especially true for dwarf galaxies, for which there is no available clustering
data. An additional implication is related to
the cumulative energy injected into the interstellar
medium in a supernova feedback-constrained scenario. Shankar et al. \citep[2006; see also][]{Dekel03}
re-derived the expected trend of stellar and halo
mass of the type $\ms \propto f_{\rm surv} \mfal^{\alpha}$. Here
$1 < \alpha < 2$ and $f_{\rm surv}$ is the fraction of surviving stellar mass.
Based on the observed rather low stellar mass in haloes with $\mfal \lesssim 10^{11}\, \msun$
inferred from a basic ABM model, they argued that supernova
feedback appeared to be insufficient to remove the gas associated with the host halo.
The results of this paper, however, show that some ABM models allow for much
larger stellar mass fraction in low mass haloes,
thus providing some hints towards the solution of this puzzling issue.

Another possible non-trivial consequence of our results concerns Halo Occupation
Distribution (HOD) models. As outlined in e.g., \citet{Berlind02}, this class
of models is based on a parameterized conditional probability $P(N|M)$ that a dark
matter halo of virial mass $M$ contains $N$ galaxies. Adding prescriptions on the spatial
distribution of subhaloes derived from accurate $N$-body simulations, HOD models have
been rather successful in reproducing the two-point correlation
function of different classes of galaxies and at different redshifts.
However, once we allow for a different ranking between satellite and central
galaxies, as was found in this paper, the shape of $P(N|M)$ might
become rather different from its usual functional form. This could induce
non-trivial degeneracies in these models,
allowing for different occupation
distributions that still match the data.

The methodology developed in this work can be used to study the clustering properties
of AGNs, cold gas, and subsets of galaxies divided by e.g. luminosity
or color. As was demonstrated here, our method already enabled us to better fit the
observed stellar mass functions and CFs of all galaxies, with
the current data sets. It might become more important when studying clustering properties of objects
which tend to depend more strongly on environment, like the cold gas mass within galaxies.
Our method may be able to shed some light on the limitations in
modeling the observed high-$\z$ galaxies using standard halo models \citep{Quadri08,Tinker10,Wake11}.
In addition, it might be useful for future surveys with high quality data and a large survey volume,
for which the simple abundance
matching approach might not be flexible enough to provide an accurate fit to the data.

Improved constraints on the $\msh$ relation might be obtained from observational
quantities, other than the stellar mass function and clustering used here.
These might include: conditional stellar mass functions per
halo mass \citep[e.g.][]{Yang09},
satellite kinematics \citep{More11}, weak lensing \citep{Mandelbaum06,Cacciato09}, star-formation histories,
and dynamical tracers such as the velocity dispersion and circular
velocity \citep[e.g.][]{Dutton10}. In addition, physically
motivated models like hydrodynamical simulations or semi-analytic models
are crucial to obtain insights on the difference between the evolution of
satellite and central galaxies. More effort to understand the fundamental relation
between the mass of subhaloes and galaxies is clearly needed both on the observational and theoretical side.


\section*{Acknowledgments}

We thank the referee for a detailed and constructive report that helped to improve the 
presentation of this work.
We also thank Peter Behroozi, Charlie Conroy, Umberto Maio, Rachel Mandelbaum,
Risa Wechsler, and Simon White for useful discussions.
The Millennium Simulation databases used in this paper and the web application providing online access to
them were constructed as part of the activities of the German Astrophysical Virtual Observatory.
FS acknowledges support from the Alexander von Humboldt Foundation.
MBK acknowledges support from the Southern California Center for Galaxy Evolution,
a multi-campus research program funded by the University of California office of 
research.

\bibliographystyle{mn2e}
\bibliography{ref_list}

\begin{thebibliography}{}

\bibitem[\protect\citeauthoryear{{Abazajian}, {Adelman-McCarthy},
  {Ag{\"u}eros}, {Allam}, {Allende Prieto}, {An}, {Anderson}, {Anderson},
  {Annis}, {Bahcall} \& et al.}{{Abazajian} et~al.}{2009}]{Abazajian09}
{Abazajian} K.~N.,  {Adelman-McCarthy} J.~K.,  {Ag{\"u}eros} M.~A.,  {Allam}
  S.~S.,  {Allende Prieto} C.,  {An} D.,  {Anderson} K.~S.~J.,  {Anderson}
  S.~F.,  {Annis} J.,  {Bahcall} N.~A.,    et al. 2009, \apjs, 182, 543

\bibitem[\protect\citeauthoryear{{Behroozi}, {Conroy} \& {Wechsler}}{{Behroozi}
  et~al.}{2010}]{Behroozi10}
{Behroozi} P.~S.,  {Conroy} C.,    {Wechsler} R.~H.,  2010, \apj, 717, 379

\bibitem[\protect\citeauthoryear{{Berlind} \& {Weinberg}}{{Berlind} \&
  {Weinberg}}{2002}]{Berlind02}
{Berlind} A.~A.,  {Weinberg} D.~H.,  2002, \apj, 575, 587

\bibitem[\protect\citeauthoryear{{Bernardi}, {Shankar}, {Hyde}, {Mei},
  {Marulli} \& {Sheth}}{{Bernardi} et~al.}{2010}]{Bernardi10}
{Bernardi} M.,  {Shankar} F.,  {Hyde} J.~B.,  {Mei} S.,  {Marulli} F.,
  {Sheth} R.~K.,  2010, \mnras, 404, 2087

\bibitem[\protect\citeauthoryear{{Binney} \& {Tremaine}}{{Binney} \&
  {Tremaine}}{1987}]{Binney87}
{Binney} J.,  {Tremaine} S.,  1987, {Galactic dynamics}

\bibitem[\protect\citeauthoryear{{Blanton} \& {Roweis}}{{Blanton} \&
  {Roweis}}{2007}]{Blanton07}
{Blanton} M.~R.,  {Roweis} S.,  2007, \aj, 133, 734

\bibitem[\protect\citeauthoryear{{Boylan-Kolchin}, {Ma} \&
  {Quataert}}{{Boylan-Kolchin} et~al.}{2008}]{Boylan08}
{Boylan-Kolchin} M.,  {Ma} C.-P.,    {Quataert} E.,  2008, \mnras, 383, 93

\bibitem[\protect\citeauthoryear{{Boylan-Kolchin}, {Springel}, {White},
  {Jenkins} \& {Lemson}}{{Boylan-Kolchin} et~al.}{2009}]{Boylan09}
{Boylan-Kolchin} M.,  {Springel} V.,  {White} S.~D.~M.,  {Jenkins} A.,
  {Lemson} G.,  2009, \mnras, 398, 1150

\bibitem[\protect\citeauthoryear{{Cacciato}, {van den Bosch}, {More}, {Li},
  {Mo} \& {Yang}}{{Cacciato} et~al.}{2009}]{Cacciato09}
{Cacciato} M.,  {van den Bosch} F.~C.,  {More} S.,  {Li} R.,  {Mo} H.~J.,
  {Yang} X.,  2009, \mnras, 394, 929

\bibitem[\protect\citeauthoryear{{Colpi}, {Mayer} \& {Governato}}{{Colpi}
  et~al.}{1999}]{Colpi99}
{Colpi} M.,  {Mayer} L.,    {Governato} F.,  1999, \apj, 525, 720

\bibitem[\protect\citeauthoryear{{Conroy} \& {Wechsler}}{{Conroy} \&
  {Wechsler}}{2009}]{Conroy09}
{Conroy} C.,  {Wechsler} R.~H.,  2009, \apj, 696, 620

\bibitem[\protect\citeauthoryear{{Conroy}, {Wechsler} \& {Kravtsov}}{{Conroy}
  et~al.}{2006}]{Conroy06}
{Conroy} C.,  {Wechsler} R.~H.,    {Kravtsov} A.~V.,  2006, \apj, 647, 201

\bibitem[\protect\citeauthoryear{{Conroy}, {Wechsler} \& {Kravtsov}}{{Conroy}
  et~al.}{2007}]{Conroy07a}
{Conroy} C.,  {Wechsler} R.~H.,    {Kravtsov} A.~V.,  2007, \apj, 668, 826

\bibitem[\protect\citeauthoryear{{Cooray} \& {Sheth}}{{Cooray} \&
  {Sheth}}{2002}]{Cooray02}
{Cooray} A.,  {Sheth} R.,  2002, \physrep, 372, 1

\bibitem[\protect\citeauthoryear{Croton et~al.,}{Croton
  et~al.}{2006}]{Croton06}
Croton D.~J.,  et~al., 2006, \mnras, 365, 11

\bibitem[\protect\citeauthoryear{{Davis}, {Efstathiou}, {Frenk} \&
  {White}}{{Davis} et~al.}{1985}]{Davis85}
{Davis} M.,  {Efstathiou} G.,  {Frenk} C.~S.,    {White} S.~D.~M.,  1985, \apj,
  292, 371

\bibitem[\protect\citeauthoryear{{De Lucia} \& {Blaizot}}{{De Lucia} \&
  {Blaizot}}{2007}]{DeLucia07}
{De Lucia} G.,  {Blaizot} J.,  2007, \mnras, 375, 2

\bibitem[\protect\citeauthoryear{{Dekel} \& {Woo}}{{Dekel} \&
  {Woo}}{2003}]{Dekel03}
{Dekel} A.,  {Woo} J.,  2003, \mnras, 344, 1131

\bibitem[\protect\citeauthoryear{{Dutton}, {Conroy}, {van den Bosch}, {Prada}
  \& {More}}{{Dutton} et~al.}{2010}]{Dutton10}
{Dutton} A.~A.,  {Conroy} C.,  {van den Bosch} F.~C.,  {Prada} F.,    {More}
  S.,  2010, \mnras, 407, 2

\bibitem[\protect\citeauthoryear{{Gao}, {Springel} \& {White}}{{Gao}
  et~al.}{2005}]{Gao05}
{Gao} L.,  {Springel} V.,    {White} S.~D.~M.,  2005, \mnras, 363, L66

\bibitem[\protect\citeauthoryear{{Guo}, {White}, {Li} \&
  {Boylan-Kolchin}}{{Guo} et~al.}{2010}]{Guo10a}
{Guo} Q.,  {White} S.,  {Li} C.,    {Boylan-Kolchin} M.,  2010, \mnras, 404,
  1111

\bibitem[\protect\citeauthoryear{{Harker}, {Cole}, {Helly}, {Frenk} \&
  {Jenkins}}{{Harker} et~al.}{2006}]{Harker06}
{Harker} G.,  {Cole} S.,  {Helly} J.,  {Frenk} C.,    {Jenkins} A.,  2006,
  \mnras, 367, 1039

\bibitem[\protect\citeauthoryear{{Hopkins}, {Croton}, {Bundy}, {Khochfar}, {van
  den Bosch}, {Somerville}, {Wetzel}, {Keres}, {Hernquist}, {Stewart},
  {Younger}, {Genel} \& {Ma}}{{Hopkins} et~al.}{2010}]{Hopkins10}
{Hopkins} P.~F.,  {Croton} D.,  {Bundy} K.,  {Khochfar} S.,  {van den Bosch}
  F.,  {Somerville} R.~S.,  {Wetzel} A.,  {Keres} D.,  {Hernquist} L.,
  {Stewart} K.,  {Younger} J.~D.,  {Genel} S.,    {Ma} C.,  2010, \apj, 724,
  915

\bibitem[\protect\citeauthoryear{Jarosik et~al.,}{Jarosik
  et~al.}{2011}]{Jarosik11}
Jarosik N.,  et~al., 2011, \apjs, 192, 14

\bibitem[\protect\citeauthoryear{{Jiang}, {Jing}, {Faltenbacher}, {Lin} \&
  {Li}}{{Jiang} et~al.}{2008}]{Jiang08}
{Jiang} C.~Y.,  {Jing} Y.~P.,  {Faltenbacher} A.,  {Lin} W.~P.,    {Li} C.,
  2008, \apj, 675, 1095

\bibitem[\protect\citeauthoryear{Kauffmann et~al.,}{Kauffmann
  et~al.}{2003}]{Kauffmann03}
Kauffmann G.,  et~al., 2003, \mnras, 341, 33

\bibitem[\protect\citeauthoryear{{Khochfar} \& {Ostriker}}{{Khochfar} \&
  {Ostriker}}{2008}]{Khochfar08}
{Khochfar} S.,  {Ostriker} J.~P.,  2008, \apj, 680, 54

\bibitem[\protect\citeauthoryear{{Khochfar} \& {Silk}}{{Khochfar} \&
  {Silk}}{2009}]{Khochfar09}
{Khochfar} S.,  {Silk} J.,  2009, \apjl, 700, L21

\bibitem[\protect\citeauthoryear{{Kravtsov}, {Berlind}, {Wechsler}, {Klypin},
  {Gottl{\"o}ber}, {Allgood} \& {Primack}}{{Kravtsov}
  et~al.}{2004}]{Kravtsov04}
{Kravtsov} A.~V.,  {Berlind} A.~A.,  {Wechsler} R.~H.,  {Klypin} A.~A.,
  {Gottl{\"o}ber} S.,  {Allgood} B.,    {Primack} J.~R.,  2004, \apj, 609, 35

\bibitem[\protect\citeauthoryear{Leauthaud et~al.,}{Leauthaud
  et~al.}{2011}]{Leauthaud11}
Leauthaud A.,  et~al., 2011, ArXiv e-prints, 1104.0928

\bibitem[\protect\citeauthoryear{{Li}, {Kauffmann}, {Jing}, {White},
  {B{\"o}rner} \& {Cheng}}{{Li} et~al.}{2006}]{Li06}
{Li} C.,  {Kauffmann} G.,  {Jing} Y.~P.,  {White} S.~D.~M.,  {B{\"o}rner} G.,
   {Cheng} F.~Z.,  2006, \mnras, 368, 21

\bibitem[\protect\citeauthoryear{{Li} \& {White}}{{Li} \& {White}}{2009}]{Li09}
{Li} C.,  {White} S.~D.~M.,  2009, \mnras, 398, 2177

\bibitem[\protect\citeauthoryear{{Mandelbaum}, {Seljak}, {Kauffmann}, {Hirata}
  \& {Brinkmann}}{{Mandelbaum} et~al.}{2006}]{Mandelbaum06}
{Mandelbaum} R.,  {Seljak} U.,  {Kauffmann} G.,  {Hirata} C.~M.,    {Brinkmann}
  J.,  2006, \mnras, 368, 715

\bibitem[\protect\citeauthoryear{{Mo}, {van den Bosch} \& {White}}{{Mo}
  et~al.}{2010}]{Mo10}
{Mo} H.,  {van den Bosch} F.~C.,    {White} S.,  2010, {Galaxy Formation and
  Evolution}

\bibitem[\protect\citeauthoryear{{Monaco}, {Murante}, {Borgani} \&
  {Fontanot}}{{Monaco} et~al.}{2006}]{Monaco06}
{Monaco} P.,  {Murante} G.,  {Borgani} S.,    {Fontanot} F.,  2006, \apjl, 652,
  L89

\bibitem[\protect\citeauthoryear{{More}, {van den Bosch}, {Cacciato}, {Skibba},
  {Mo} \& {Yang}}{{More} et~al.}{2011}]{More11}
{More} S.,  {van den Bosch} F.~C.,  {Cacciato} M.,  {Skibba} R.,  {Mo} H.~J.,
   {Yang} X.,  2011, \mnras, 410, 210

\bibitem[\protect\citeauthoryear{{Moster}, {Somerville}, {Maulbetsch}, {van den
  Bosch}, {Macci{\`o}}, {Naab} \& {Oser}}{{Moster} et~al.}{2010}]{Moster10}
{Moster} B.~P.,  {Somerville} R.~S.,  {Maulbetsch} C.,  {van den Bosch} F.~C.,
  {Macci{\`o}} A.~V.,  {Naab} T.,    {Oser} L.,  2010, \apj, 710, 903

\bibitem[\protect\citeauthoryear{{Neistein} \& {Weinmann}}{{Neistein} \&
  {Weinmann}}{2010}]{Neistein10}
{Neistein} E.,  {Weinmann} S.~M.,  2010, \mnras, 405, 2717

\bibitem[\protect\citeauthoryear{{Neistein}, {Weinmann}, {Li} \&
  {Boylan-Kolchin}}{{Neistein} et~al.}{2010}]{Neistein11}
{Neistein} E.,  {Weinmann} S.~M.,  {Li} C.,    {Boylan-Kolchin} M.,  2010,
  ArXiv e-prints: 1011.2492 (paper I)

\bibitem[\protect\citeauthoryear{{Pasquali}, {van den Bosch}, {Mo}, {Yang} \&
  {Somerville}}{{Pasquali} et~al.}{2009}]{Pasquali09}
{Pasquali} A.,  {van den Bosch} F.~C.,  {Mo} H.~J.,  {Yang} X.,    {Somerville}
  R.,  2009, \mnras, 394, 38

\bibitem[\protect\citeauthoryear{{Purcell}, {Bullock} \& {Zentner}}{{Purcell}
  et~al.}{2007}]{Purcell07}
{Purcell} C.~W.,  {Bullock} J.~S.,    {Zentner} A.~R.,  2007, \apj, 666, 20

\bibitem[\protect\citeauthoryear{{Quadri}, {Williams}, {Lee}, {Franx}, {van
  Dokkum} \& {Brammer}}{{Quadri} et~al.}{2008}]{Quadri08}
{Quadri} R.~F.,  {Williams} R.~J.,  {Lee} K.,  {Franx} M.,  {van Dokkum} P.,
  {Brammer} G.~B.,  2008, \apjl, 685, L1

\bibitem[\protect\citeauthoryear{{Sawala}, {Guo}, {Scannapieco}, {Jenkins} \&
  {White}}{{Sawala} et~al.}{2011}]{Sawala11}
{Sawala} T.,  {Guo} Q.,  {Scannapieco} C.,  {Jenkins} A.,    {White} S.,  2011,
  \mnras, pp 64--+

\bibitem[\protect\citeauthoryear{{Shankar}, {Lapi}, {Salucci}, {De Zotti} \&
  {Danese}}{{Shankar} et~al.}{2006}]{Shankar06}
{Shankar} F.,  {Lapi} A.,  {Salucci} P.,  {De Zotti} G.,    {Danese} L.,  2006,
  \apj, 643, 14

\bibitem[\protect\citeauthoryear{{Skibba} \& {Sheth}}{{Skibba} \&
  {Sheth}}{2009}]{Skibba09}
{Skibba} R.~A.,  {Sheth} R.~K.,  2009, \mnras, 392, 1080

\bibitem[\protect\citeauthoryear{{Skibba}, {van den Bosch}, {Yang}, {More},
  {Mo} \& {Fontanot}}{{Skibba} et~al.}{2011}]{Skibba11}
{Skibba} R.~A.,  {van den Bosch} F.~C.,  {Yang} X.,  {More} S.,  {Mo} H.,
  {Fontanot} F.,  2011, \mnras, 410, 417

\bibitem[\protect\citeauthoryear{{Springel}, {White}, {Jenkins}, {Frenk},
  {Yoshida}, {Gao}, {Navarro}, {Thacker}, {Croton}, {Helly}, {Peacock}, {Cole},
  {Thomas}, {Couchman}, {Evrard}, {Colberg} \& {Pearce}}{{Springel}
  et~al.}{2005}]{Springel05}
{Springel} V.,  {White} S.~D.~M.,  {Jenkins} A.,  {Frenk} C.~S.,  {Yoshida} N.,
   {Gao} L.,  {Navarro} J.,  {Thacker} R.,  {Croton} D.,  {Helly} J.,
  {Peacock} J.~A.,  {Cole} S.,  {Thomas} P.,  {Couchman} H.,  {Evrard} A.,
  {Colberg} J.,    {Pearce} F.,  2005, \nat, 435, 629

\bibitem[\protect\citeauthoryear{{Springel}, {White}, {Tormen} \&
  {Kauffmann}}{{Springel} et~al.}{2001}]{Springel01}
{Springel} V.,  {White} S.~D.~M.,  {Tormen} G.,    {Kauffmann} G.,  2001,
  \mnras, 328, 726

\bibitem[\protect\citeauthoryear{{Tinker}, {Wechsler} \& {Zheng}}{{Tinker}
  et~al.}{2010}]{Tinker10}
{Tinker} J.~L.,  {Wechsler} R.~H.,    {Zheng} Z.,  2010, \apj, 709, 67

\bibitem[\protect\citeauthoryear{{Tinker}, {Weinberg}, {Zheng} \&
  {Zehavi}}{{Tinker} et~al.}{2005}]{Tinker05}
{Tinker} J.~L.,  {Weinberg} D.~H.,  {Zheng} Z.,    {Zehavi} I.,  2005, \apj,
  631, 41

\bibitem[\protect\citeauthoryear{{Vale} \& {Ostriker}}{{Vale} \&
  {Ostriker}}{2004}]{Vale04}
{Vale} A.,  {Ostriker} J.~P.,  2004, \mnras, 353, 189

\bibitem[\protect\citeauthoryear{{van den Bosch}, {Tormen} \& {Giocoli}}{{van
  den Bosch} et~al.}{2005}]{vdBosch05}
{van den Bosch} F.~C.,  {Tormen} G.,    {Giocoli} C.,  2005, \mnras, 359, 1029

\bibitem[\protect\citeauthoryear{{von der Linden}, {Best}, {Kauffmann} \&
  {White}}{{von der Linden} et~al.}{2007}]{vdLinden07}
{von der Linden} A.,  {Best} P.~N.,  {Kauffmann} G.,    {White} S.~D.~M.,
  2007, \mnras, 379, 867

\bibitem[\protect\citeauthoryear{Wake et~al.,}{Wake  et~al.}{2011}]{Wake11}
Wake D.~A.,  et~al., 2011, \apj, 728, 46

\bibitem[\protect\citeauthoryear{{Wang}, {Li}, {Kauffmann} \& {De
  Lucia}}{{Wang} et~al.}{2006}]{Wang06}
{Wang} L.,  {Li} C.,  {Kauffmann} G.,    {De Lucia} G.,  2006, \mnras, 371, 537

\bibitem[\protect\citeauthoryear{{Wang}, {Li}, {Kauffmann} \& {De
  Lucia}}{{Wang} et~al.}{2007}]{Wang07}
{Wang} L.,  {Li} C.,  {Kauffmann} G.,    {De Lucia} G.,  2007, \mnras, 377,
  1419

\bibitem[\protect\citeauthoryear{{Weinmann}, {Kauffmann}, {von der Linden} \&
  {De Lucia}}{{Weinmann} et~al.}{2010}]{Weinmann10}
{Weinmann} S.~M.,  {Kauffmann} G.,  {von der Linden} A.,    {De Lucia} G.,
  2010, \mnras, 406, 2249

\bibitem[\protect\citeauthoryear{{Weinmann}, {van den Bosch}, {Yang} \&
  {Mo}}{{Weinmann} et~al.}{2006}]{Weinmann06}
{Weinmann} S.~M.,  {van den Bosch} F.~C.,  {Yang} X.,    {Mo} H.~J.,  2006,
  \mnras, 366, 2

\bibitem[\protect\citeauthoryear{{Weinmann}, {van den Bosch}, {Yang}, {Mo},
  {Croton} \& {Moore}}{{Weinmann} et~al.}{2006}]{Weinmann06a}
{Weinmann} S.~M.,  {van den Bosch} F.~C.,  {Yang} X.,  {Mo} H.~J.,  {Croton}
  D.~J.,    {Moore} B.,  2006, \mnras, 372, 1161

\bibitem[\protect\citeauthoryear{{Yang}, {Mo} \& {van den Bosch}}{{Yang}
  et~al.}{2009a}]{Yang09}
{Yang} X.,  {Mo} H.~J.,    {van den Bosch} F.~C.,  2009a, \apj, 695, 900

\bibitem[\protect\citeauthoryear{{Yang}, {Mo} \& {van den Bosch}}{{Yang}
  et~al.}{2009b}]{Yang09a}
{Yang} X.,  {Mo} H.~J.,    {van den Bosch} F.~C.,  2009b, \apj, 693, 830

\bibitem[\protect\citeauthoryear{{Yang}, {Mo}, {van den Bosch}, {Pasquali},
  {Li} \& {Barden}}{{Yang} et~al.}{2007}]{Yang07}
{Yang} X.,  {Mo} H.~J.,  {van den Bosch} F.~C.,  {Pasquali} A.,  {Li} C.,
  {Barden} M.,  2007, \apj, 671, 153

\bibitem[\protect\citeauthoryear{York et~al.,}{York  et~al.}{2000}]{York00}
York D.~G.,  et~al., 2000, \aj, 120, 1579

\bibitem[\protect\citeauthoryear{Zehavi et~al.,}{Zehavi
  et~al.}{2005}]{Zehavi05}
Zehavi I.,  et~al., 2005, \apj, 630, 1

\bibitem[\protect\citeauthoryear{{Zentner}, {Berlind}, {Bullock}, {Kravtsov} \&
  {Wechsler}}{{Zentner} et~al.}{2005}]{Zentner05}
{Zentner} A.~R.,  {Berlind} A.~A.,  {Bullock} J.~S.,  {Kravtsov} A.~V.,
  {Wechsler} R.~H.,  2005, \apj, 624, 505

\end{thebibliography}

\appendix
\section{The location of satellite galaxies}
\label{sec:sat_loc}

Here we describe the model we use for fixing the location of unresolved subhaloes. These subhaloes
are not identified at $\z=0$, but can still host galaxies according to the dynamical friction estimate.

Consider a massive point particle subject to dynamical friction in an infinite
mass distribution with density profile $\rho \propto r^{-\gamma}$.  In the limit
of a circular orbit for the point particle and an isothermal profile
($\gamma=2$) for the background density, the decay of the satellite's orbit due
to dynamical friction can be modeled analytically (eq. 7-25 of Binney \&
Tremaine 1987).  Assuming that $\gamma$ does not differ much from 2,
we can follow the same arguments as Binney and Tremaine and
find that the equation governing the decay rate for the specific angular momentum $L$ is
\begin{equation}
  \label{eq:a0}
\frac{\dd L}{\dd t} = \frac{F\,r}{M_{\rm sat}} \approx -c \ln\Lambda \,M_{\rm sat} \,r\, \frac{\rho(r)}{v_c^2(r)} \,,
\end{equation}
where $\ln\Lambda$ is the Coulomb logarithm (assumed to be constant), $c$ is a constant,
$v_c$ is the circular velocity, and $F$ is the frictional force. Using $L=rv_c$,
and $v_c \propto r^{-\gamma/2+1}$ we get
\begin{equation}
\label{eq:a1}
\frac{\dd L}{\dd t} \propto \frac{\dd r}{\dd t}\,r^{-\gamma/2+1} \propto -\,r^{-1}  \,.
\end{equation}
Integrating, we obtain
\begin{equation}
r=r_{\rm sat}(1-\tau)^{1/(3-\gamma/2)} \,,
\end{equation}
where $\tau$ is the fraction of the initially estimated dynamical friction
time-scale that has elapsed until $\z=0$, i.e. $t(\z=0)-t(\z_{\rm sat}) = \tau
\,t_{\rm df}$, and $\z_{\rm sat}$ is the last redshift the subhalo was identified.
For additional flexibility, we allow $\tau$ to vary as well,
corresponding to the overall uncertainty in the dynamical friction time-scale:
\begin{equation}
\label{eq:a2}
r=r_{\rm sat}(1-\tau^p)^{1/q}\,.
\end{equation}

\begin{figure}
\centerline{ \hbox{ \epsfig{file=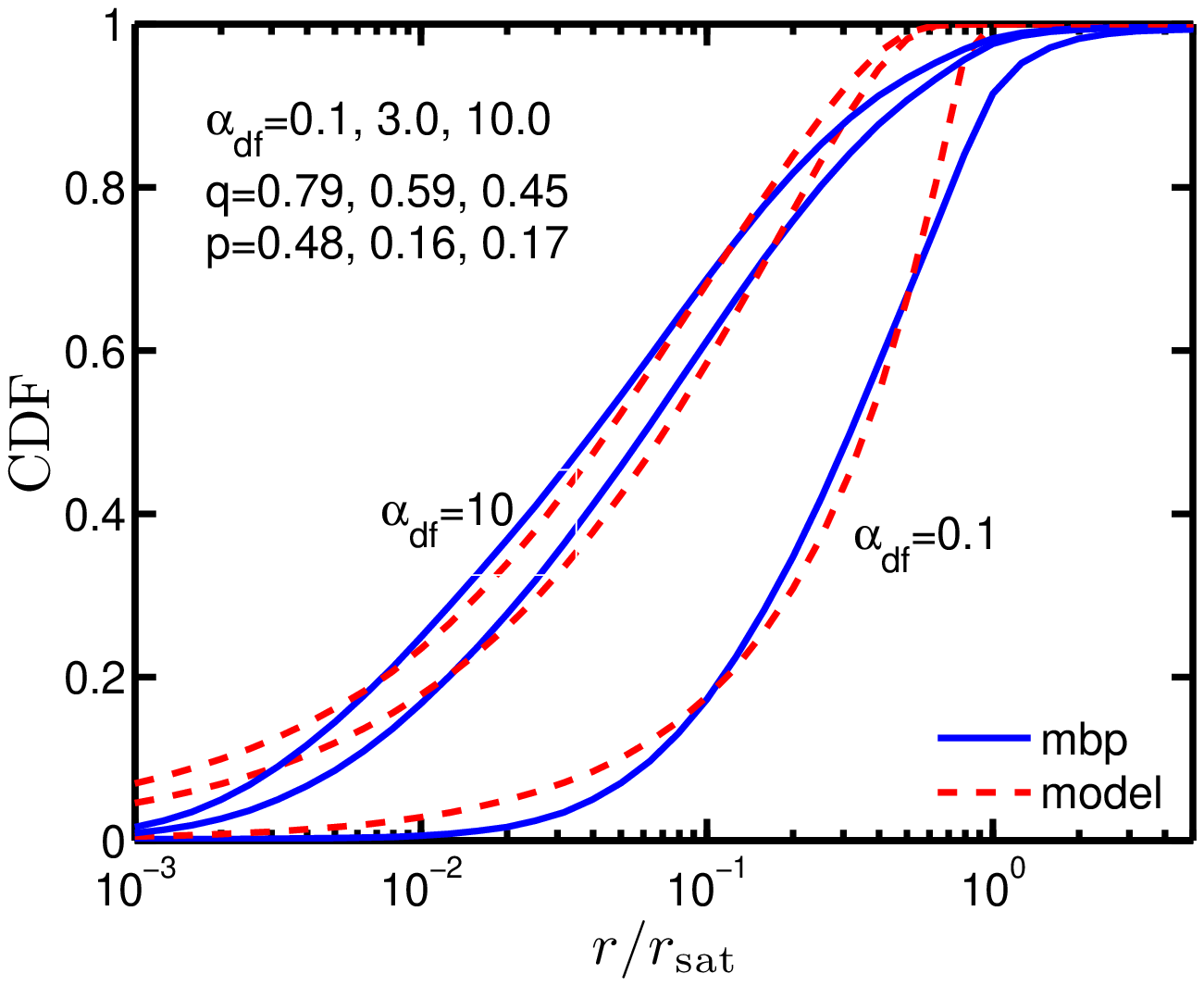,width=9cm} }}
\caption{The cumulative distribution function (CDF) of $r/r_{\rm sat}$. CDF using
the location of the most bound particle are plotted
in \emph{solid lines} for $\adf=0.1,3,10$ as indicated. For each value of $\adf$ we plot
in \emph{dashed line} the best fitting model using Eq.~\ref{eq:a2}.}
  \label{fig:mbp_model}
\end{figure}

In Fig.~\ref{fig:mbp_model} we compare the distribution of $r/r_{\rm sat}$ when using the most bound particle, against the model
summarized in Eq.~\ref{eq:a2}.
Subhaloes from the simulation are chosen according to $\adf$. For each unresolved subhalo we compute the location
of the last identified most bound particle, this is shown in solid lines. Dashed lines show the results when
using the model discussed above.
For each value of $\adf$ plotted, we have optimized the parameters $p,q$ to get the best fit. Typically,
$r_{\rm sat}$ is lower than 1 Mpc, and the CF measurements are reliable above $\sim10$ kpc. Consequently, the difference between
the plotted lines at $r/r_{\rm sat}<10^{-2}$ is of less importance to this study. We have checked that the values of $p,q$ plotted here, agree with
the location set by the most bound particle also when computing the CFs. Lastly, we list in Table \ref{tab:models} the combinations of satellite
locations and $\adf$ that were used for the full parameter search discussed in section \ref{sec:results}.

\begin{table}
\caption{The set of parameters used for scanning the parameter space in section \ref{sec:results}. `mbp' refers to a model
with satellite locations set by the most bound particle of the last identified subhalo.}
\begin{center}
\begin{tabular}{lccc}
\hline Satellite location (p) & q & & $\adf$  \\
\hline
mbp  &  --  & & 0.1, 3, 8, 10 \\
0.1  &  0.4 & & 0.1, 3     \\
0.14 & 0.6 & & 0.1, 3, 10     \\
0.5 & 0.8 & & 0.1, 1, 3, 10     \\
1.0 &  1.0 & & 0.1, 1     \\
0.17 & 0.4 & & 10     \\
\hline
\end{tabular}
\end{center}
\label{tab:models}
\end{table}
%

\section{Scanning the parameter space}
\label{sec:scan_prm}

As explained in section \ref{sec:results}, each model within our formalism is defined using 18 free
parameters that fix the five domains in stellar mass used here. There are additional three parameters
that define the dynamical friction time scale ($\adf$), and the location of unresolved subhaloes ($p,q$).
Even though this seems to be a huge parameter space, we managed to scan it systematically and with high resolution.
In this section we explain the numerical details that allow this parameter search.

\begin{itemize}
  \item Searching the parameter space is done separately for the
  different domains of stellar mass. This reduces the number of independent parameters
  to those that define a domain (i.e. 6). Only in a later stage we search
  for solutions that combine a set of 5 domains that have mutual boundaries.
  \item We search for solutions one domain after another, starting from the most massive one. At each domain
we require that the model boundaries will coincide with an accepted model from the previous domain.
  \item We first check that the stellar mass function meets the
  fitting criterion. Only after we have a solution for all the domains
  we check the fit to the CF.
  \item Due to the number of bins we adopt for $\mfal$ and $\delta_i$
each domain has $10^{10}$ optional combinations of $\alpha_i,\beta_i,\delta_i$.
We first compute separately the contributions
to the mass function (Eq.~\ref{eq:N}) from satellite and central subhaloes ($\phi_c,\, \phi_s$). 
Only in a later stage we combine all
the $10^{10}$ options to see which of them fits the data.
  \item Before computing the CF we sort the different models such that there will be a minimum
difference between neighboring models in the list. We then compute the CF using the `moving
average' scheme. 
This means that we do not compute the integral of Eq.~\ref{eq:N} (and the related
integral involving $\psi$) for each model, but
only compute the incremental changes of this integral when going over the full list of models.
  \item We compute CF values starting the smallest radius. After the CF is computed for each radius, we
check whether the model fits the data. This allows us to quickly reject bad solutions.
\end{itemize}
Our algorithm is able to compute CF for each domain within a
$10^{-2}$ second, using one processor. In total, we have computed
the CF for more than $\sim10^7$ models. The CFs for all models are based on the full
sample of subhaloes from the Millennium simulation.

\label{lastpage}

\end{document}